\documentclass[%
 aip,
 amsmath,amssymb,
 reprint,%
]{revtex4-1}

\usepackage{afterpage}
\usepackage{amsmath}
\usepackage{amssymb}
\usepackage{amsfonts}
\usepackage{amsthm}

\newcommand{\be}{\begin{equation}}
\newcommand{\ee}{\end{equation}}
\newcommand{\bea}{\begin{eqnarray}}
\newcommand{\eea}{\end{eqnarray}}

\newcommand{\bs}[1]{\boldsymbol{#1}}

\usepackage{graphicx}
\usepackage{dcolumn}
\usepackage{bm}

\usepackage[utf8]{inputenc}
\usepackage[T1]{fontenc}
\usepackage{mathptmx}

\begin{document}

\preprint{AIP/123-QED}

\title{Microwave Hall effect measurement for materials in the skin depth region}

\author{Ryo Ogawa}
\affiliation{%
Department of Basic Science, the University of Tokyo, 3-8-1 Komaba, Meguro-ku, Tokyo, 153-8902, Japan
}%

\author{Tatsunori Okada}
\affiliation{
Institute for Materials Research, Tohoku University, 2-1-1 Katahira, Aoba-ku, Sendai, Miyagi, 980-8577, Japan
}%
\author{Hideyuki Takahashi}
\affiliation{
Molecular Photoscience Research Center, Kobe University, 1-1 Rokkodai-cho, Nada, Kobe, 657-8501,  Japan
}%
\author{Fuyuki Nabeshima}%
\author{Atsutaka Maeda}
\affiliation{%
Department of Basic Science, the University of Tokyo, 3-8-1 Komaba, Meguro-ku, Tokyo, 153-8902, Japan
}%

\date{\today}
\begin{abstract}
We developed a new  microwave Hall effect measurement method for materials in the skin depth region at low temperatures using a cross-shaped bimodal cavity.
We analytically calculated electromagnetic fields in the cross-shaped cavity, and the response of the cavity including the sample, whose property is represented by the surface impedance tensor; further, we constructed the method to obtain the Hall component of the surface impedance tensor in terms of the change in resonance characteristics.
To confirm the validity of the new method, we applied our method to measure the Hall effect in metallic Bi single crystals at low temperatures, and we confirmed that the microwave Hall angles coincide with the DC Hall angle.
Thus, it becomes clear that the Hall angle measurement under cryogenic conditions becomes possible without any complicated tuning mechanisms, and our bimodal cavity method can be used to measure the microwave Hall effect on materials in the skin depth region.
The result opens a new approach to discuss the Hall effect in condensed matter physics such as the microwave flux-flow Hall effect in superconductors. 
\end{abstract}

\maketitle

\section{Introduction}

The Hall effect is one of the most effective techniques to understand the mechanism of electric conduction in materials.
To measure the Hall effect, electrodes are attached perpendicular to the direction of the flow of current. 
However, in cases such as those involving powder samples and very narrow quasi-one dimensional materials, 
it is difficult to realize such an experimental configuration; further, in other cases, nondestructive measurements are required.
For these cases, noncontact  methods that use high-frequency, in particular, microwave electromagnetic fields have been utilized.~\cite{Cooke1948,Portis1958,Liu1961,Nishina1961,Watanabe1961,Sayed1975a,Sayed1975,Ong1981}

Thus far, several studies have investigated the microwave Hall effect meaurement.
\cite{Cooke1948,Portis1958,Sayed1975a,Sayed1975,Ong1977,Nishina1958,Hambleton1959,Liu1961,Nishina1961,Watanabe1961,Fletcher1976,Ong1981,Kuchar1986,Na1992,Prati2003,Murthy2006,Cross1980,Eley1983,Al-Zoubi2005,Cross1980,Eley1983,Al-Zoubi2005,Coppock2016} 
In these studies, the most commonly used approach is the use of a bimodal cavity, which is a resonator with two orthogonal degenerate modes.\cite{Cooke1948,Portis1958}
Since the two orthogonal degenerate modes are coupled with each other through the Hall effect, the bimodal cavity can be used to measure the Hall effect.
In Table~\ref{table}, we summarize the previous representative measurements for the microwave Hall effect using  bimodal cavities.
As listed in Table~\ref{table}, many studies investigated semiconductors, insulators,~\cite{Portis1958,Nishina1958,Hambleton1959,Liu1961,Nishina1961,Watanabe1961,Sayed1975a,Sayed1975,Fletcher1976,Ong1977,Ong1981,Kuchar1986,Na1992,Prati2003,Murthy2006} or biomaterials,\cite{Cross1980,Eley1983,Al-Zoubi2005} where the conductivity is rather low, thereby making
the Hall effect measurement easy.
Ong et al. proposed an improved method for materials with slightly higher conductivity.~\cite{Ong1981}
However, there is lack of studies that investigate the microwave Hall measurement in the skin depth regime, i.e., in highly conductive materials.

To improve sensitivity and accuracy, apparatuses used in these studies employ some complicated mechanisms to tune the mode of the cavity.~\cite{Portis1958,Nishina1958,Nishina1961,Watanabe1961,Sayed1975,Fletcher1976,Ong1977,Cross1980,Ong1981,Na1992,LinfengChen1998,Prati2003}
However, under cryogenic conditions, it is difficult to employ these mechanisms.
Therefore, most previous studies were conducted at room temperature, which is indicated in Table~\ref{table}.

Since much attention has been given to understanding the dynamics of the quantized magnetic vortex in superconductors,\cite{Kopnin1995,Blatter1994,Kopnin2002} 
there is an urgent need to study the AC Hall effect in highly conductive materials under  cryogenic conditions.
Thus, we develop a new microwave Hall effect measurement method for materials in the skin depth region at low temperatures. 

In this paper, we developed a new microwave Hall effect measurement method for materials in the skin depth region using a cross-shaped bimodal cavity.
Our method can be applied under a wide range of temperatures, including cryogenic conditions.
We analytically calculated the electromagnetic fields in the cross-shaped cavity  and the response of the cavity including the sample, whose property is represented by the surface impedance tensor; further, we developed the method to obtain the Hall component of the surface impedance tensor in terms of the change in resonance characteristics.

To confirm the validity of the proposed method, we applied our method to measure the Hall effect in metallic Bi single crystals at low temperatures, and we confirmed that the microwave Hall angle coincides with the DC Hall angle. 
Thus, Hall angle measurement under cryogenic conditions is possible without the use of any complicated tuning mechanisms such as screws as long as the tangent of the Hall angle is on the order of 0.1.
The result presents a new approach to discuss the Hall effect in condensed matter physics.
In particular, it is now possible to measure the Hall effect in superconductors with a finite magnetic field at sufficiently lower temperatures than the superconducting transition temperature, which could not be investigated by the DC measurement.

The rest of this paper is organized as follows: 
In the next section, we introduce newly derived formulas for our experimental configuration, which are necessary to obtain the Hall angle; further, we explain how to obtain the microwave Hall angle.
The details of the calculation are described in the Appendix. 
In Sections. III and IV, we describe the experimental details and measurement results for the metallic Bi single  crystals.
Finally, the paper is concluded in Section V with a summary of our results.

\begin{table*}[htbp]\label{table}
\begin{minipage}{\hsize}
\centering
\caption{Previous studies of the microwave Hall effect by bimodal cavities. Cyl., Rect. and Crossed W. G. are abbreviations for cylindrical, rectangular and crossed waveguide, respectively.}
\begin{tabular}{cccrrrrcc} \hline\hline
\multicolumn{1}{c}{1st Auther} & \multicolumn{1}{c}{Year} & \multicolumn{1}{c}{Shape} & \multicolumn{1}{c}{$f$ (GHz)} &\multicolumn{1}{c}{$Q$}&\multicolumn{1}{c}{$T$ (K)}&\multicolumn{1}{c}{$B$(T)}&\multicolumn{1}{c}{Materials}&\multicolumn{1}{c}{Ref.} \\ \hline
	S. Cooke & 1948 & Cyl. & 9& ---&R.T.&0.3&Metals&\cite{Cooke1948} \\ 
	E. Portis & 1958 & Cyl. & X Band& ---&R.T.& ---&CuSO$_4$$\cdot$5H$_2$O&\cite{Portis1958} \\ 
	G. Hambleton & 1959 & Crossed W.G. & 20& ---&80&1 &Si, Ge&\cite{Hambleton1959} \\ 
	Y. Nishina & 1961 & Rect. &9&2000&30&0.2&Ge&\cite{Nishina1961,Liu1961} \\ 
	N. Watanabe & 1961 & Cyl. &24& ---&100&0.5&Ge&\cite{Watanabe1961} \\ 
	M. Sayed & 1975 & Cyl. &9&2200&R.T.&0.7&CdS&\cite{Sayed1975a,Sayed1975} \\ 
	J. Fletcher & 1976 & Rect. &X band&6500&R.T.&1&ZnO&\cite{Fletcher1976} \\ 
	N. Ong & 1977 & Cyl. &9& ---&R.T.&1.2&TTF-TCNQ&\cite{Ong1977} \\ 
	T. Cross & 1980 & Cyl. &33&15000&R.T.& ---&Biopolymers&\cite{Cross1980} \\ 
	N. Ong & 1981 & Cyl. &9&3000&R.T.& ---&Si&\cite{Ong1981} \\ 
	D. Eley & 1983 & Cyl. &9&9500&R.T.&1.2&Biological samples&\cite{Eley1983} \\ 
	F. Kuchar & 1986 & Crossed W.G. &33& ---&2.2&8&GaAs-AlGaAs&\cite{Kuchar1986} \\ 
	M. Dressel & 1991 &  Cyl. &10& ---&R.T.&0.8&BEDT-TTF, YBCO&\cite{Dressel1991} \\ 
	B. Na & 1992 &  Cyl. and Rect. &X band&4000&R.T.&0.5&ZnO&\cite{Na1992} \\ 
	L. Chen & 1998 &  Cyl.&X band&2000&R.T.&0.3&Iron oxide &\cite{LinfengChen1998} \\ 
	E. Prati & 2003 &  Cyl. &20& ---&R.T.&0.6&GaAs, ZnSe&\cite{Prati2003} \\ 
	A. Al-Zoubi & 2005 &  Cyl. and Rect. &X band&6000&R.T.& ---&Organic semiconductors&\cite{Al-Zoubi2005} \\ 
	D. Murthy& 2006 &  Cyl. &10&3000&R.T.&0.6&Si, InSb&\cite{Murthy2006} \\ 
	D. Murthy& 2008 &  Cyl. &14&3100&R.T.&0.6&C-nanotube&\cite{Murthy2008} \\ \hline
\end{tabular}
\end{minipage}
\end{table*}

\section{Method}

The Hall angle $\theta$ is represented as
\be
\tan\theta=\frac{\sigma_{xy}}{\sigma_{xx}},
\ee
where $\sigma_{xx}$ and $\sigma_{xy}$ are the $xx$ and $xy$ components of the conductivity tensor in the $xy$ plane $\tilde{\sigma}$, which is defined as
\be
\bs{j}=\tilde{\sigma}\bs{E},
\ee
where $\bs{j}\equiv(j_x, j_y)$ and $\bs{E}\equiv(E_x, E_y)$ are the current density and electric field vectors, respectively.
We assumed that the magnetic field $\bs{B}$ is in the $z$ direction.
Then, we rewrite the conductivity tensor using $\tan\theta$ as
\begin{eqnarray}\label{appendix_simga}
	\tilde{\sigma}
	&=&\sigma_{xx}
	\left(
	\begin{array}{cc}
		1 &-\tan\theta\\
		\tan\theta&1 \\
	\end{array}
	\right).
\end{eqnarray}
Note that $\tan\theta$ defined here is different from that in DC measurement; it is a complex quantity.
When the single carrier model is appropriate, $\tan\theta$ is revised to $\tan\theta_{dc}/(1-i\omega\tau)$, where $\omega$ denotes the angular frequency, and $\tau$ denotes the scattering time of the carrier.

As shown in Eqs.~(\ref{appendix_zsigmagannma}) and (\ref{appendix_z}), the surface impedance tensor  in the $xy$ plane is
\be
\tilde{Z}\equiv \tilde{R} - i\tilde{X}  = \left(
\begin{array}{cc}
Z^L & Z^H \\
-Z^H & Z^L \\
\end{array}
\right),
\ee 
and it is related to $\tilde{\sigma}$ as
\be
\tilde{Z}=\gamma\tilde{\sigma}^{-1},
\ee
where $\gamma$ denotes the inverse of the skin depth, and $\tilde{R}$ and $\tilde{X}$ denote the real  (surface resistance) and  imaginary parts (surface reactance) of $\tilde{Z}$.
The explicit expression of $\gamma$ is not necessary for the discussions provided below.
As shown in the Appendix (Eq.~(\ref{eq:appendix_ZHZ})), the Hall angle is expressed by the ratio of the components in the surface impedance tensor as 
\be
\label{method_tan}
\tan\theta=\frac{Z^H}{Z^L}.
\ee

We consider a cross-shaped bimodal cavity and analytically calculate the electromagnetic fields in the cavity, where bimodal modes TE$_{101}$ and TE$_{011}$ exist at the resonance frequency $f_{H}$ with the quality factor $Q_H$.
We derive the relationship between the resonance characteristics and surface impedance tensor, and we express $\tan\theta$ in terms of the surface impedance tensor (See Appendix for details on the derivation).
As a result, we found that  changes in the resonant frequency and $Q$ factor of the resonance  are related to the components of the surface impedance tensor as
\begin{equation}\label{method_crossQ}
\Delta_{w/wo}\left(\frac{1}{2Q_H}\right)\equiv\frac{1}{2Q_H}-\frac{1}{2Q_{H0}}=G^LR^L+G^H|X^H|
\end{equation}
and
\begin{equation}\label{method_crossf}
\Delta_{w/wo}\left(\frac{f_H}{f_{H0}}\right)\equiv -\frac{f_{H}-f_{H0}}{f_{H0}}=G^LX^L-G^H|R^H|+D,
\end{equation}
where $\Delta_{w/wo}$ represents the difference between the data with and without the sample; $f_H$ and $f_{H0}$ are the resonance frequencies with and without the sample, respectively; $Q$ and $Q_0$ are the $Q$ factor of the resonance with and without the sample, respectively; $R^L$ and $X^L$ are the longitudinal components of the surface resistance and surface reactance tensors, respectively; $R^H$ and $X^H$ are the off-diagonal components of the surface resistance and surface reactance tensors, respectively; $G^L$ and $G^H$ are the geometrical constants in the longitudinal and Hall directions of the bimodal cavity, respectively, which are the same magnitude for the ideally symmetrical case; 
and $D$ is the experimentally inevitable constant.
When the Hall effect is absent, these formulas are equivalent to equations for an ordinary cylindrical cavity perturbation technique.\cite{Klein1993,Donovan1993,Dressel1993}
These formulas where the real and imaginary parts of the surface impedance tensor are intermingled can be interpreted in terms of the adiabatic theorem which states that the shift in complex frequency is equal to the change in the complex energy of the cavity. 
The real part of the complex energy corresponds to the energy stored in the cavity, and the imaginary part corresponds to energy loss. 
The mixing of the real and imaginary parts of the system corresponds to the energy exchange between the orthogonal modes caused by the Hall effect. 
In other words, it implies that the energy loss in one mode is transferred to the energy of the other mode by the Hall effect. 

Below, we show an explicit method to obtain $\tan\theta$ from the above presented new formulas. 
Eq.~(\ref{method_tan}) shows that if we obtain $R^L$, $X^L$, $R^H$ and $X^H$, we can find $\tan\theta$. 
Hence, we perform three types of measurements; (1) DC resistivity measurement, (2)  ordinary surface impedance measurement using a cylindrical cavity resonator, and (3) Hall surface impedance measurement using the cross-shaped bimodal cavity.

First,  we determine longitudinal components of the surface impedance tensor $Z^L= R^L - i X^L$, 
which can be obtained with the ordinary cavity perturbation technique using a cylindrical cavity; here, the shift of the resonant frequency and $Q$ factor of the resonance are related to $Z^L$ as
\begin{equation}\label{method_cylindricalQ}
\Delta_{w/wo}\left(\frac{1}{2Q}\right)\equiv\frac{1}{2Q}-\frac{1}{2Q_0}=GR^L
\end{equation}
and
\begin{equation}\label{method_cylindricalf}
\Delta_{w/wo}\left(\frac{f}{f_0}\right)\equiv -\frac{f-f_0}{f_0}=GX^L+C,
\end{equation}
where $f$ and $f_0$ denote the resonance frequencies with and without the sample, respectively; $Q$ and $Q_0$ denote  the $Q$ factors of the resonance with and without the sample, respectively; $C$ 
denotes the experimentally inevitable constant; and $G$ denotes the geometrical constant.
Note that $Q$ and $Q_H$ or $f$ and $f_H$ are quantities in different resonators.
Further, $G$ and $C$ were determined using the assumption that $R^L=X^L=(\mu_0\omega/2\sigma_{dc})^{1/2}$, where $\mu_0$, $\omega=2\pi f$, and $\sigma_{dc}$ are the vacuum permeability, angular frequency, and DC conductivity, respectively.
Thus, we need to measure  the longitudinal components of DC resistivity.

Next, the measurement are performed using the bimodal cavity. 
Changes in the resonance characteristics of the cross-shaped bimodal cavity are related to the surface impedance tensor of the sample as indicated in Eqs.~(\ref{method_crossQ}) and (\ref{method_crossf}).
Therefore, we need to determine $G^L$, $G^H$, and $D$, which are independent of temperature, to obtain the off-diagonal component of the surface impedance tensor $Z^H$. 
 In the cross-shaped cavity measurement of this study, we delete $D$ by considering the difference between the data of the same sample at different temperatures rather than $\Delta_{w/wo}$.
Thus,  we rewrite Eqs.~(\ref{method_crossQ}) and (\ref{method_crossf}) as
\begin{equation}\label{method_DcrossQ}
\Delta\left(\frac{1}{2Q_H}\right)(T:T_0, B)=G^L\Delta R^L(T:T_0, B)+G^H\Delta |X^H(T:T_0, B)|
\end{equation}
and
\begin{equation}\label{method_Dcrossf}
\Delta\left(\frac{f_H}{f_{H0}}\right)(T:T_0, B)=G^L\Delta X^L(T:T_0, B)-G^H\Delta |R^H(T:T_0, B)|,
\end{equation}
where $\Delta$ is defined by
\begin{equation}\label{method_Delta}
\Delta A(T:T_0,B)=A(T,B)-A(T_0,B),
\end{equation}
where $A$ is some physical quantity such as the surface resistance.

To  determine geometrical factors $G^L$ and $G^H$, we measure the magnetic field dependence of the responses $Q_H$ and $f_H$ at two temperatures $T_0$ and $T_1$, and we evaluate $\Delta(1/2Q_H)(T_1:T_0,B)$ and $\Delta(f_H/f_{H0})(T_1:T_0,B)$.
Then, we consider two assumptions: (1) the tangents of the DC Hall angles are small at $T_0$ and $T_1$, and (2) the DC Hall angle and microwave Hall angle have the same magnitude at $T_0$ and $T_1$.
We confirm that assumption (1) is valid for Bi in the results section.
The assumption (2) is valid at low frequencies such as microwaves provided the relaxation time(s) of the carrier(s) is(are) not as large as that in the anomalous skin region
These assumptions are valid for many materials  including high T$_{c}$ superconductors in the normal state.

From assumption (1), considering that $G^L$ and $G^H$ are the same order of magnitude, for Bi, we get
\begin{equation}
|X^H|-|R^H|\sim|\tan\theta_{dc}|(|X^L|-|R^L|)\ll R^L+X^L,
\end{equation}
where $\tan\theta_{dc}$ is the tangent of the Hall angle in the DC measurement.
Thus, by taking the sum of Eqs.~(\ref{method_DcrossQ}) and (\ref{method_Dcrossf}), we approximate
\begin{eqnarray}\label{method_approx1}
&&\Delta\left(\frac{1}{2Q_H}\right)(T_1:T_0, B)+\Delta\left(\frac{f_H}{f_{H0}}\right)(T_1:T_0, B)\nonumber \\
&&\simeq G^L[\Delta R^L(T_1:T_0, B)+\Delta X^L(T_1:T_0, B)].
\end{eqnarray}
From Eq.~(\ref{method_approx1}), $G^L$ is determined with the data of $Q_H$ and $f_H$, and the surface impedance data $R^L$ and $X^L$ at $T_1$ and $T_{0}$.

\begin{figure}[htbp]
		\includegraphics[keepaspectratio,width=80mm]{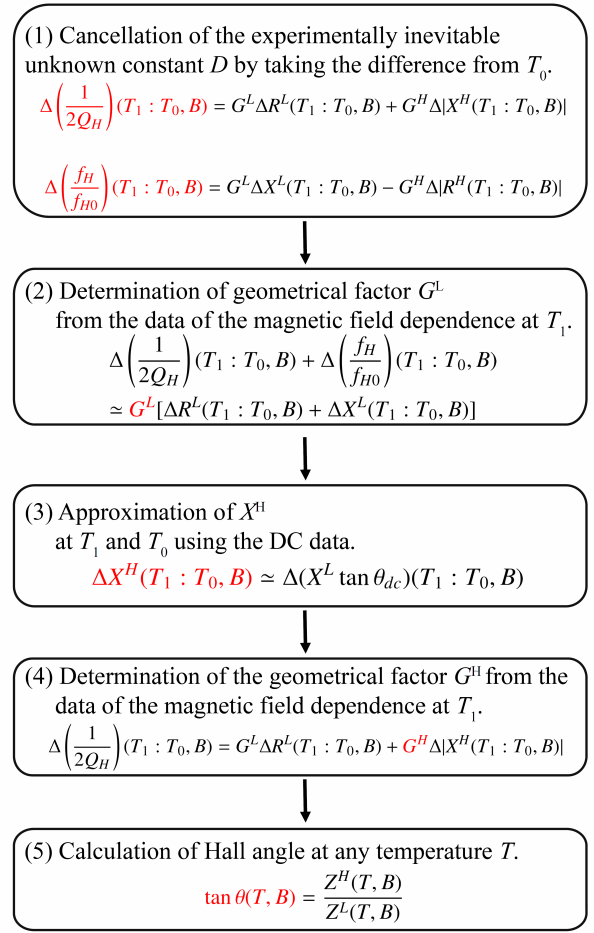}
		\caption{Flowchart of the analytic procedure for the obtained resonance characteristics from cross-shaped cavity measurement in Bi. 
		In each equation, the physical quantity obtained in that procedure is represented in red. The black letters represent the physical quantities already known at the time.}
		\label{flowchart}
\end{figure}

From assumption (2), we obtain
$\Delta R^H(T_1:T_0, B)\simeq \Delta(R^L\tan\theta_{dc})(T_1:T_0, B)$ or $\Delta X^H(T_1:T_0, B)\simeq \Delta(X^L\tan\theta_{dc})(T_1:T_0, B)$.
Then, $G^H$ is obtained from Eq.~(\ref{method_DcrossQ}) or (\ref{method_Dcrossf}) together with the data of $Q_H$ and $f_H$ and the longitudinal components of the surface impedance tensor at $T_{1}$ and $T_{0}$. 
Thus, in this method for Bi, we need to measure the Hall component of DC resistivity at $T_{1}$ and $T_{0}$.
Once the geometrical factors are determined, we can obtain $R^H$ and $X^H$.
Considering the ratio of $Z^L$ to $Z^H$, we can obtain the tangent of the microwave Hall angle at any temperature without the DC resistivity measurement at the temperature. 
It is necessary to determine the sign of the Hall angle using other measurements (e.g., DC resistivity measurements).
We summarize the procedure in Fig.~\ref{flowchart}.

In addition, there are several approaches to determine $G^L$. 
One of the approaches is as follows.
We measure the temperature dependence of the responses in a zero magnetic field $\Delta(1/2Q_H)(T:T_0, 0)$ and $\Delta (f_H/f_{H0})(T:T_0, 0)$, where the Hall terms in the formulas vanish. 
Thus, $G^L$ is obtained from $\Delta(1/2Q_H)(T:T_0, 0)=G^L\Delta R^L(T:T_0, 0)$ or $\Delta (f_H/f_{H0})(T:T_0, 0)=G^L\Delta X^L(T:T_0, 0)$ with the surface impedance data already obtained in the measurement using the cylindrical cavity. 
This approach is effective for materials with a large change in surface impedance caused by temperature change, such as superconductors.
However, for the case of metallic Bi, the above mentioned approach is more effective because of the intensity of the signal.

To summarize, let us repeat the explicit method to obtain $\tan\theta$.
(1) First, the DC resistivity is measured. (2) Next, the longitudinal component of the surface impedance tensor $Z^L$ is measured by the cylindrical cavity.
(3) Finally, the off-diagonal component of the surface impedance tensor $Z^H$ is measured by the cross-shaped bimodal cavity.
With this procedure, the microwave Hall effect can be measured at any temperature by measuring the DC resistivity tensor at a few temperatures.

Indeed, we can apply another method of analysis using our formulas to determine the geometrical factors from a sample of known properties, and then, we can investigate the microwave Hall effect for another sample of similar shape and unknown properties using the  determined geometrical factors.
With this method, the microwave Hall effect can be measured without the DC measurement of the sample to be studied.
However, we selected the previously mentioned procedure because it allows us to analyze the properties of samples without considering the effects of subtle differences in the shape of the sample.

\section{Experiments}
For the measurement, we chose metallic Bi because of its large DC Hall angle and high conductivity.
Bi single crystals were prepared by melting Bi polycrystals above 271.4 $^\circ$C, post cooling, and cleaving of the precipitated columnar crystals along the (111) plane according to the method discussed in Ref~\cite{Yamamoto1951}. 
Single crystals were cut into rectangular pieces with typical dimensions of 0.8$\times$0.8$\times$0.1 mm$^3$.
Measurements were performed using three samples (\#S1, \#S2, and \#S3), and we obtained essentially the same results in all three samples.
Therefore, we show the representative results of only one sample, \#S1, below.

The DC measurements were performed using a standard six-probe configuration using a Quantum Design PPMS (model 6000). 
The longitudinal component of $\tilde{Z}$ was measured by the cylindrical cavity perturbation technique.
We used a cylindrical oxygen-free Cu cavity resonator operated in the TE$_{011}$ mode at 19 GHz, which has a quality factor $Q\sim 6\times10^4$.

\begin{figure}[htbp]
		\includegraphics[keepaspectratio,width=70mm]{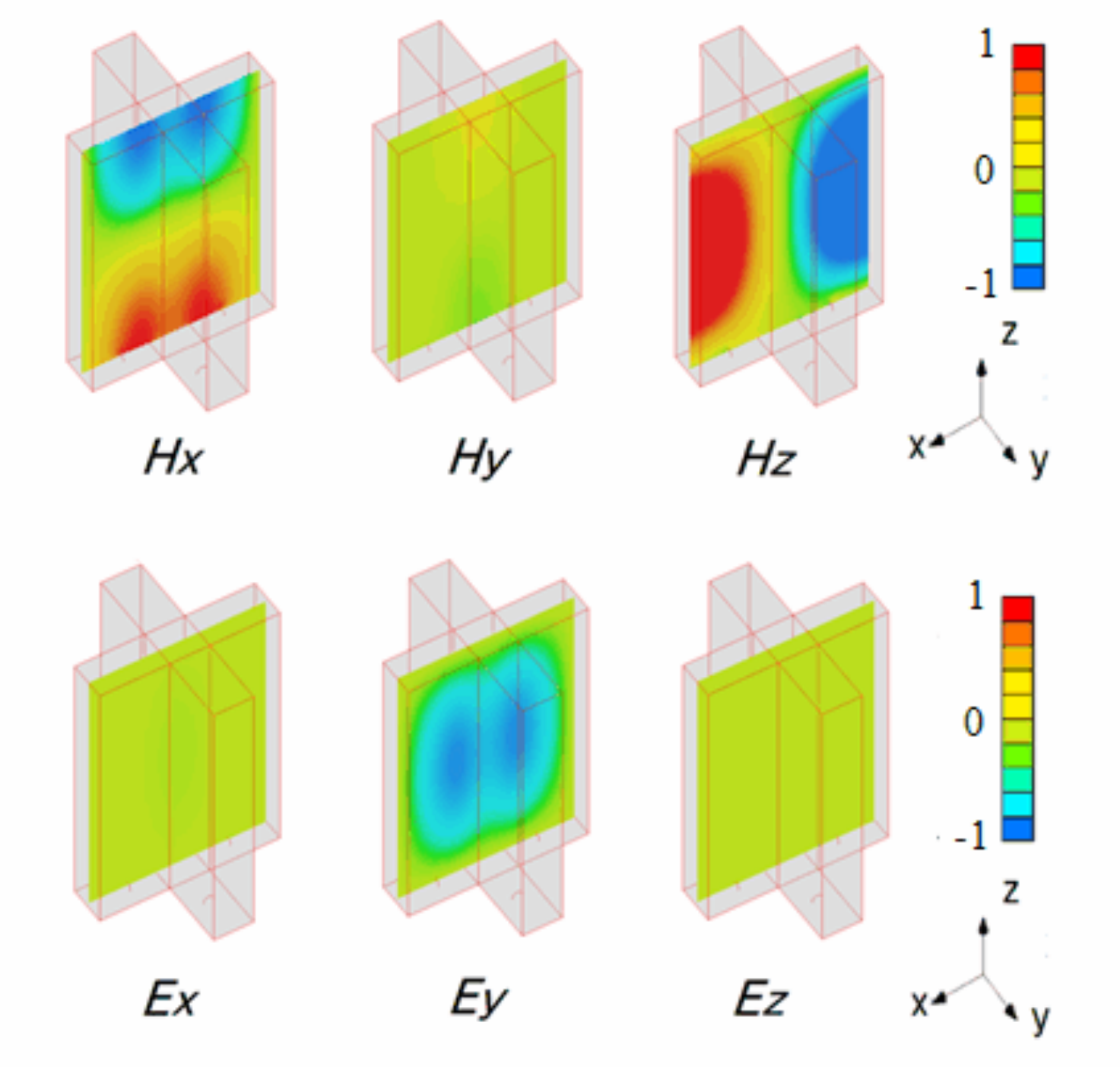}
		\caption{Simulation of the  electromagnetic field in the cross-shaped bimodal cavity at 15.8 GHz, where the TE$_{101}$ and TE$_{011}$ modes exist.
		The normalized electric field $\bm{E}$ and magnetic field $\bm{H}$ in the $xz$ plane are shown.
		Red and blue color represent the positive and negative directions, respectively.
		The electric field is oriented in the $y$ axis and the magnetic field rotates in the $xz$ plane.
		That is, it shows the existence of the TE$_{101}$ mode.}
		\label{simulation}
\end{figure}

\begin{figure}[htbp]
		\includegraphics[keepaspectratio,width=60mm]{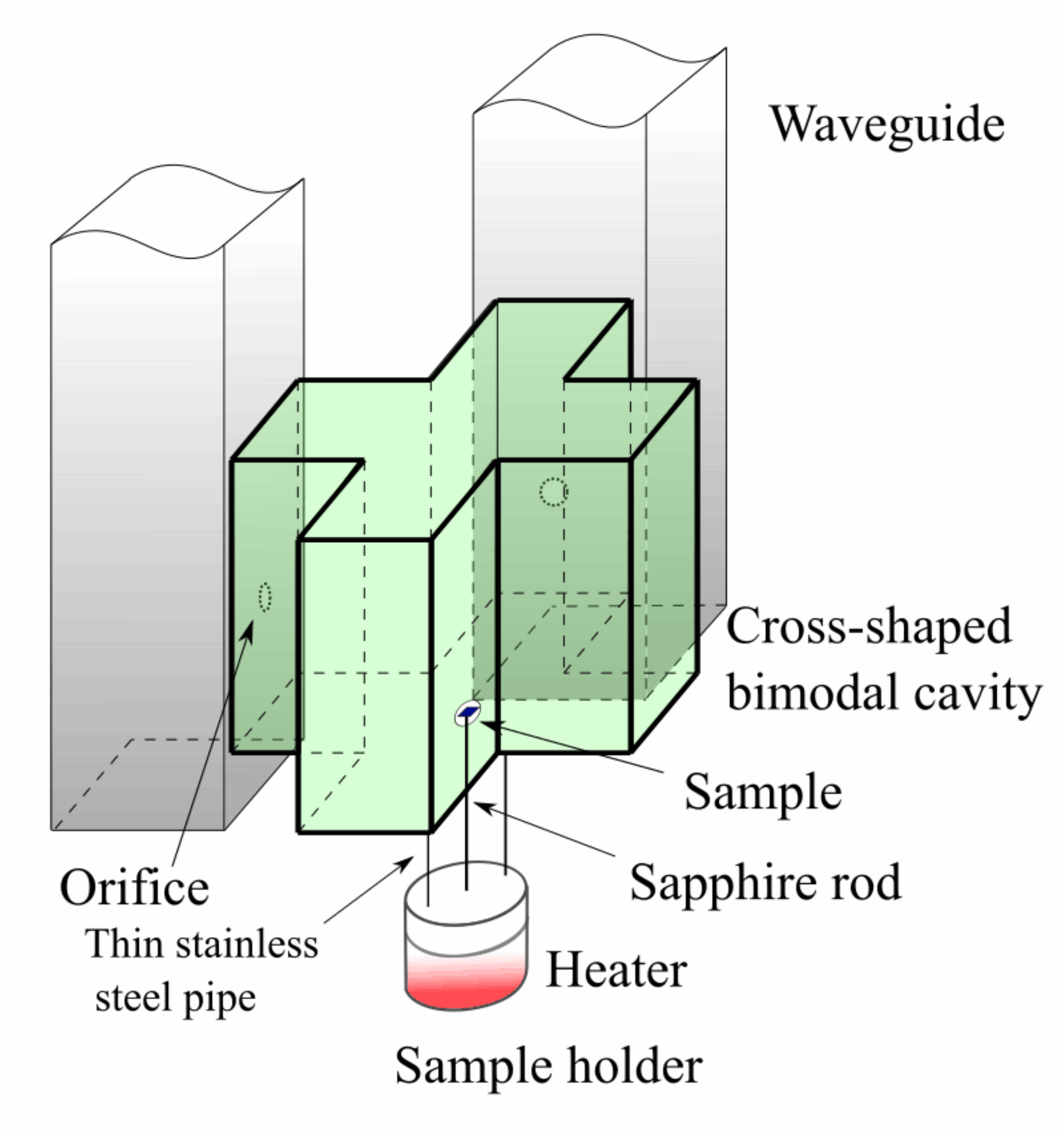}
		\caption{Sketch of the cross-shaped bimodal cavity.}
		\label{cavity}
\end{figure}

\begin{figure}[htbp]
		\includegraphics[keepaspectratio,width=70mm]{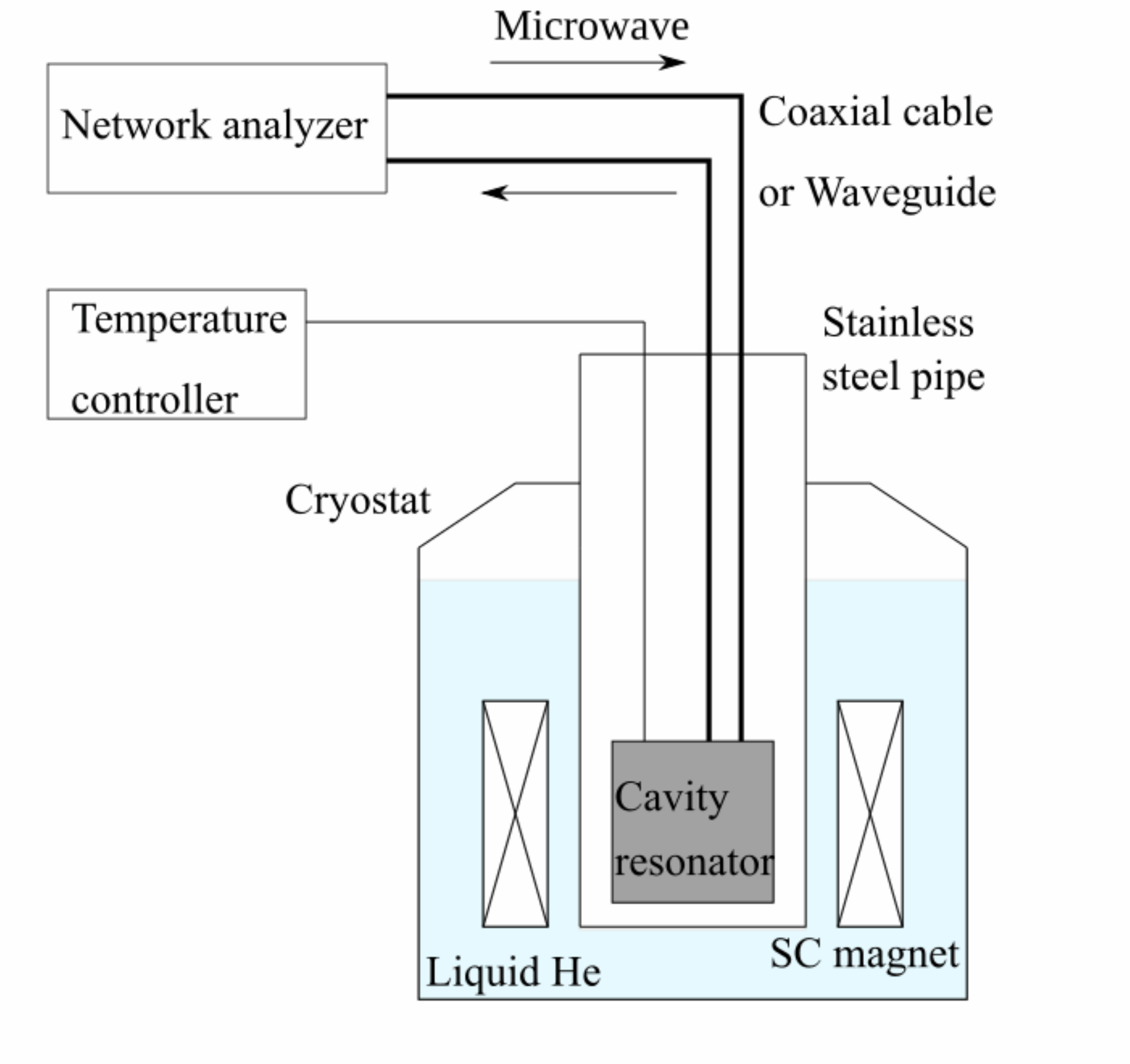}
		\caption{Overall sketch of the cross-shaped cavity measurement system.}
		\label{system}
\end{figure}

The cross-shaped bimodal cavity for the microwave Hall effect measurement was fabricated from oxygen-free Cu. 
Waveguides were attached at its protruded parts, and they were connected to the cavity through an orifice of about 1 mm diameter. 
Each length of the long side of the cross was 10.5 mm, and its height was 54 mm; it was operated at 15.8 GHz, where the two orthogonal degenerate modes such as TE$_{011}$ and TE$_{101}$ in a rectangular resonator exist.
The excitation of these bimodal modes near 15.8 GHz was confirmed by the numerical simulation of the  electromagnetic fields shown in Fig.~\ref{simulation}.
In the real cross-shaped bimodal cavity, these modes have a quality factor $Q\sim 3\times10^3$. 
A network analyzer (Agilent Technologies N5222A) was used to measure the resonant frequency and $Q$ factor of the resonance.
The cross-shaped bimodal cavity was placed in an evacuated pipe made of stainless steel, which is submerged in liquid He for cryogenic measurements. 
The cavity walls are connected with the sample holder containing the sapphire rod and heater only through different cooled thin stainless steel pipes. 
Therefore, the sapphire rod and the cavity wall are thermally well isolated. 
Because the sample is placed on the sapphire rod, we can change the temperature of the sample alone without changing the temperature of the cavity (the “hot-finger” method\cite{Sridhar1988}).
Thus, we can consider that $G^L$, $G^H$, and $D$ are independent of temperature.
The overall sketch of our cross-shaped cavity measurement system is shown in Figs.~\ref{cavity} and \ref{system}.


\section{Results}

\begin{figure}[htbp]
		\begin{minipage}[t]{\hsize}
			\mbox{\includegraphics[keepaspectratio,width=70mm]{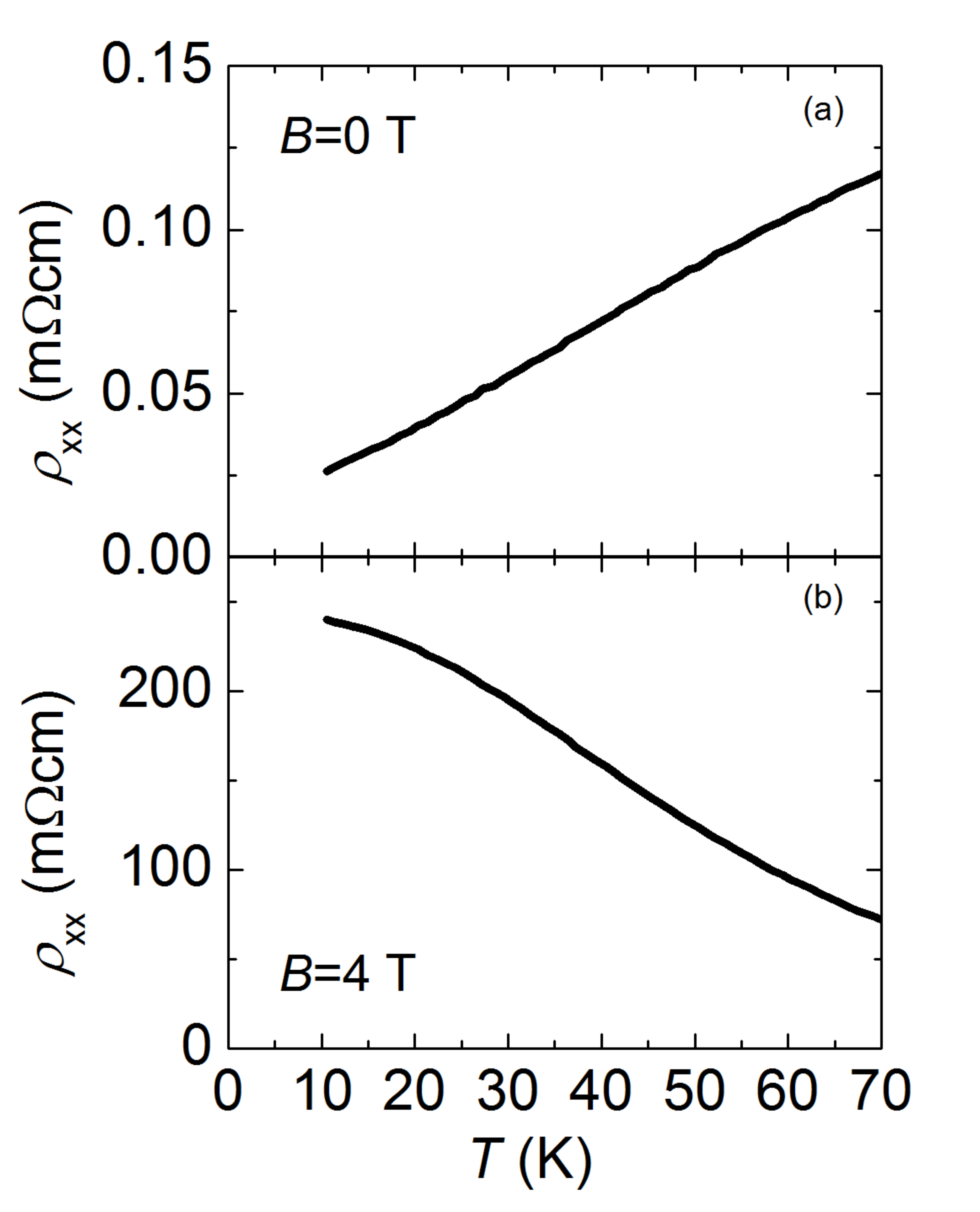}} 
 		\end{minipage}
		\begin{minipage}[t]{\hsize}
			\mbox{\includegraphics[keepaspectratio,width=70mm]{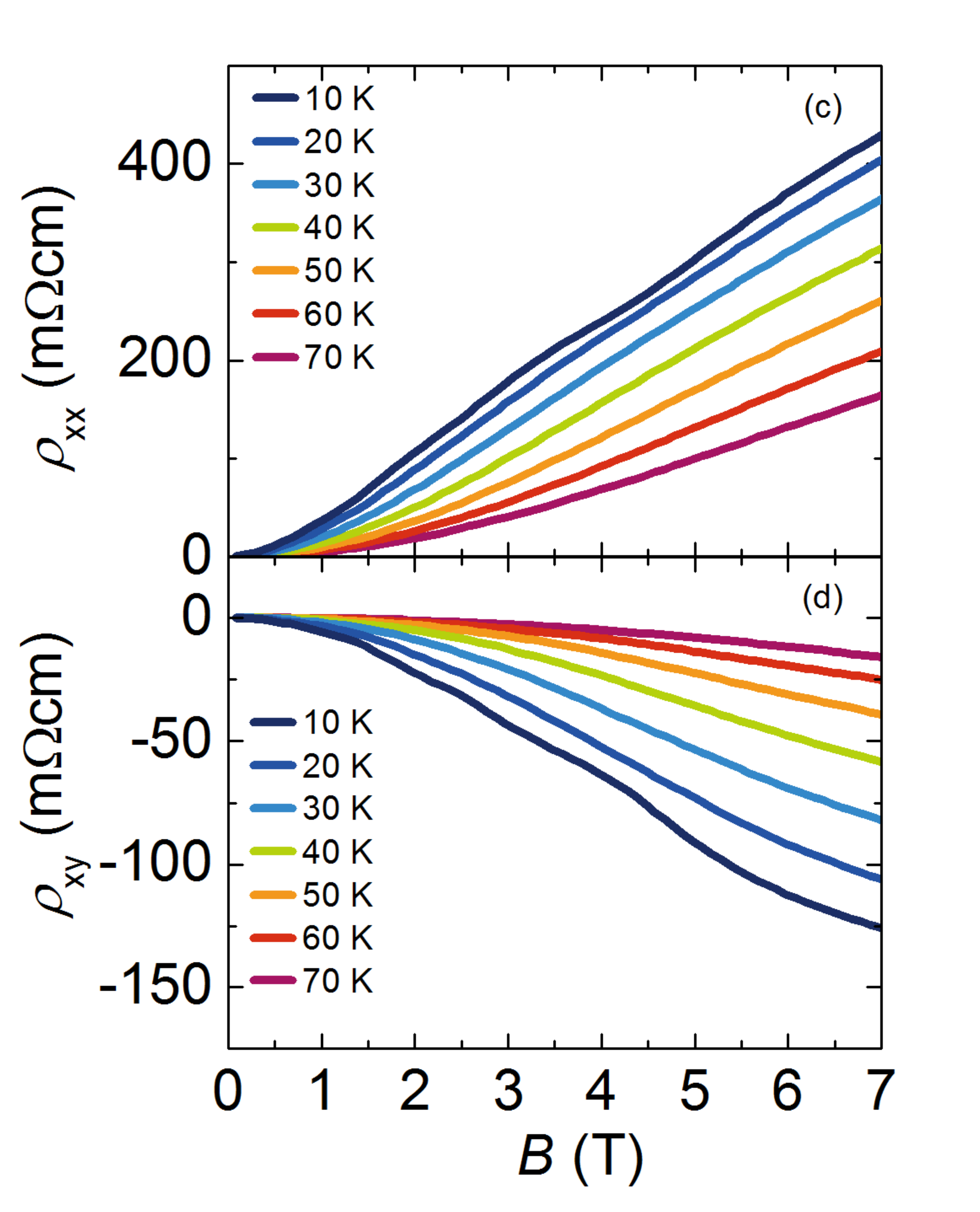}} 
		\end{minipage}
	\caption{Longitudinal resistivity $\rho_{xx}$ of Bi sample \#S1 as a function of temperature at (a) $B=0$ T and (b) $B=4$ T. In a magnetic field, the resistivity becomes large at low temperatures. (c) The longitudinal resistivity $\rho_{xx}$ and (d) transverse resistivity $\rho_{xy}$ of \#S1 as a function of the magnetic field. The Shubnikov-de Haas oscillation can be seen at low temperatures especially at 10 K.}
	\label{DC}
\end{figure}

Figures~\ref{DC}(a) and (b) show the temperature dependence of the DC longitudinal resistivity $\rho_{xx}$ at $B=0$ T and $B=4$ T. 
The magnetic field is applied perpendicular to the (111) plane in all  measurements shown below.
Figure~\ref{DC}(b) shows that, in a magnetic field, $\rho_{xx}$ becomes large at low temperatures. 
In Figs.~\ref{DC}(c) and (d), the DC longitudinal ($\rho_{xx}$) and transverse resistivity ($\rho_{xy}$) are shown as a function of the magnetic field.
We can see the Shubnikov-de Haas oscillation at low temperatures, particularly at 10 K.
The tangent of the Hall angle at 7 T reaches -0.3 at 10 K and -0.1 at 60 K.
These behaviors are essentially the same as those reported previously in the literature, \cite{Alers1953,Abeles1956}which show the high quality of our crystal.

\begin{figure}[htbp]
	\mbox{\includegraphics[keepaspectratio,width=70mm]{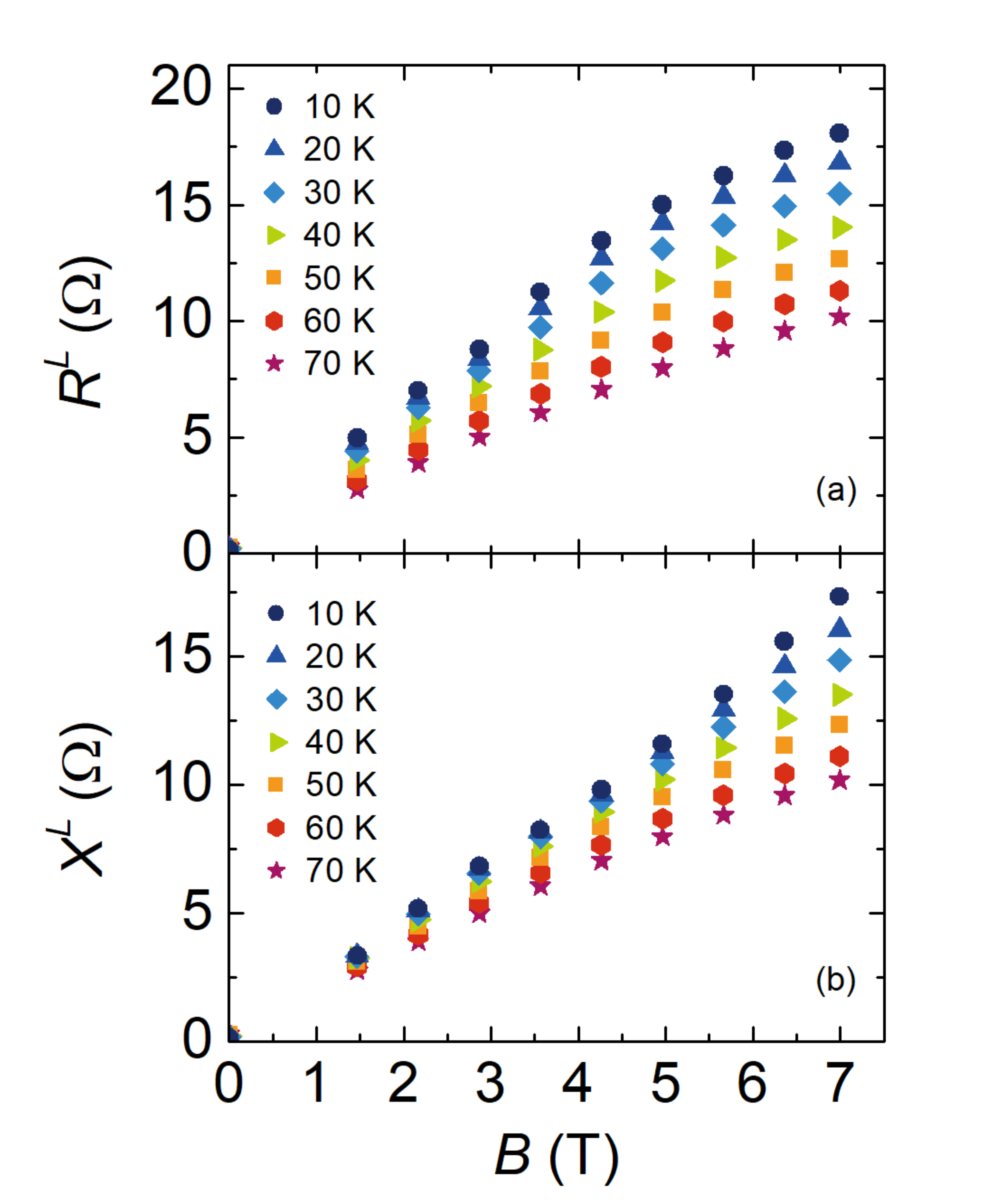}} 
	\caption{(a) Surface resistance and (b) surface reactance of \#S1 as a function of magnetic field.
	These data were obtained by the cylindrical cavity perturbation technique.}
	\label{Cyl}
\end{figure}

Figures \ref{Cyl}(a) and (b) illustrate the magnetic field dependence of the longitudinal components of the surface impedance tensor $R^L$ and $X^L$, respectively.
These data were obtained by the cylindrical cavity perturbation technique.
As expected from the DC measurements, $R^L$ and $X^L$ increase with an increasing magnetic field and with deceasing temperature in the magnetic fields. 

\begin{figure}[htbp]
	\includegraphics[keepaspectratio,width=70mm]{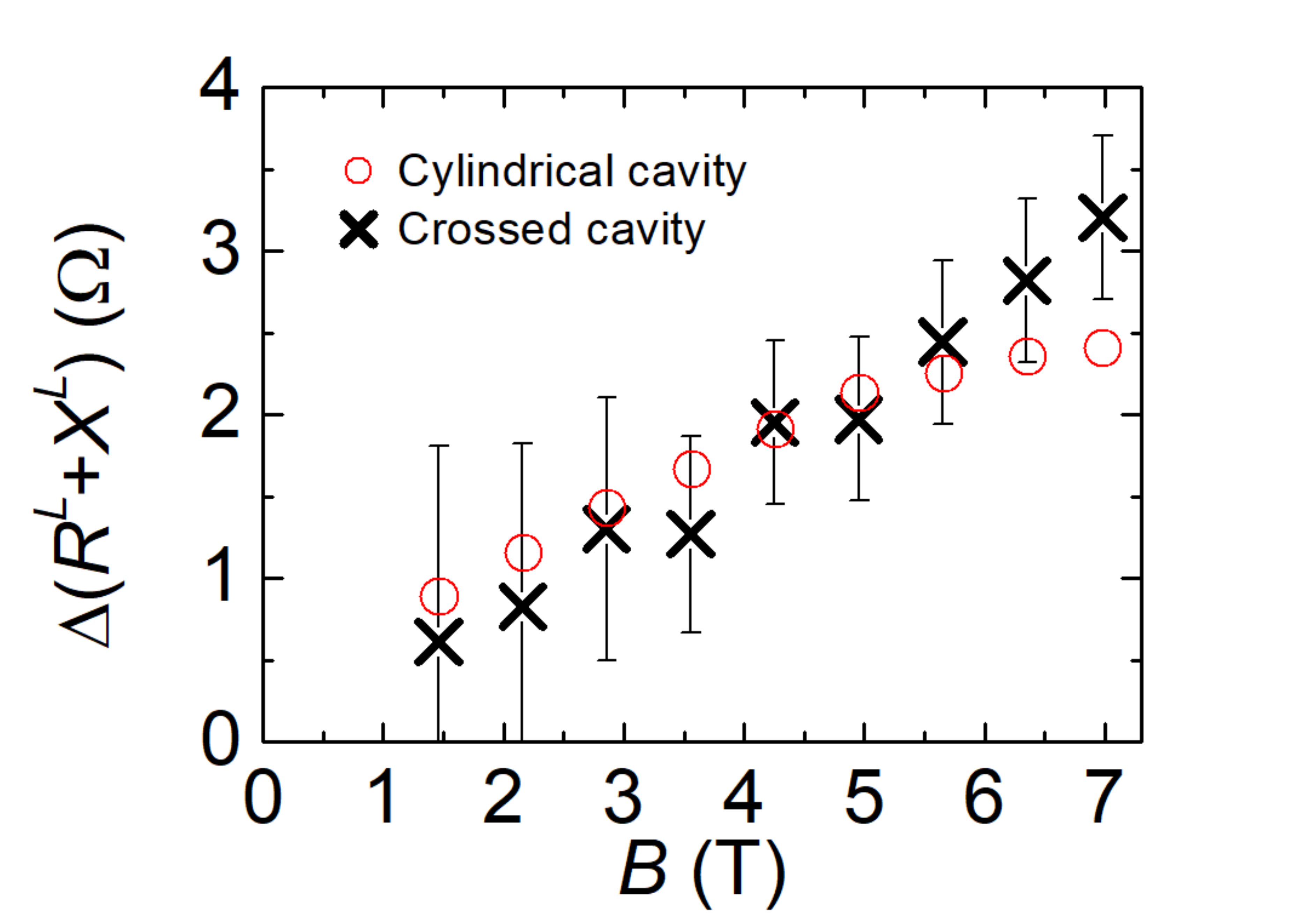}
	\caption{Magnetic field dependence of $\Delta R^L(T_1:T_0,B)+\Delta X^L(T_1:T_0,B)$ of \#S1. 
	Red circles: Data obtained by cylindrical cavity measurements. 
	Cross marks: $[\Delta(1/2Q_H)(T_1:T_0,B)+\Delta(f_H/f_{H0})(T_1:T_0,B)]/G^L$ data obtained by cross-shaped cavity measurements. 
	The geometric factor $G^L$ was set to $G^L=5.6\pm 0.8\times 10^{-8}$ $\Omega^{-1}$.}
	\label{GL}
\end{figure}

In the bimodal cavity measurement, we first determine $G^L$ and $G^H$.
From Eqs.~(\ref{method_DcrossQ})-(\ref{method_approx1}) in Section II, we compare the data of $\Delta(1/2Q_H)(T_1:T_0,B)+\Delta(f_H/f_{H0})(T_1:T_0,B)$ measured in the crossed cavity with the $\Delta R^L(T_1:T_0,B)+\Delta X^L(T_1:T_0,B)$ data measured in the cylindrical cavity. 
As for the difference $\Delta$, we consider  $T_0$ and $T_1$ as 70 K and 60 K, respectively.
We selected the high temperatures of 60 K and 70 K to avoid the influence of the anomalous skin effect.
After performing the cross-shaped cavity measurement together with the surface impedance data obtained in the cylindrical cavity shown in Fig. \ref{Cyl}, the geometric factor $G^L$ was obtained as $G^L=5.6\pm 0.8\times 10^{-8}$ $\Omega^{-1}$ from Eq.~(\ref{method_approx1}).
With this $G^L$, both the $[\Delta(1/2Q_H)(T_1:T_0,B)+\Delta(f_H/f_{H0})(T_1:T_0,B)]/G^L$ data and $\Delta R^L(T_1:T_0,B)+\Delta X^L(T_1:T_0,B)$ data are shown in a comparative manner  in Fig.~\ref{GL}.
The error bars represent the noise of signals and systematic errors of measurement systems.

\begin{figure}[htbp]
	\includegraphics[keepaspectratio,width=70mm]{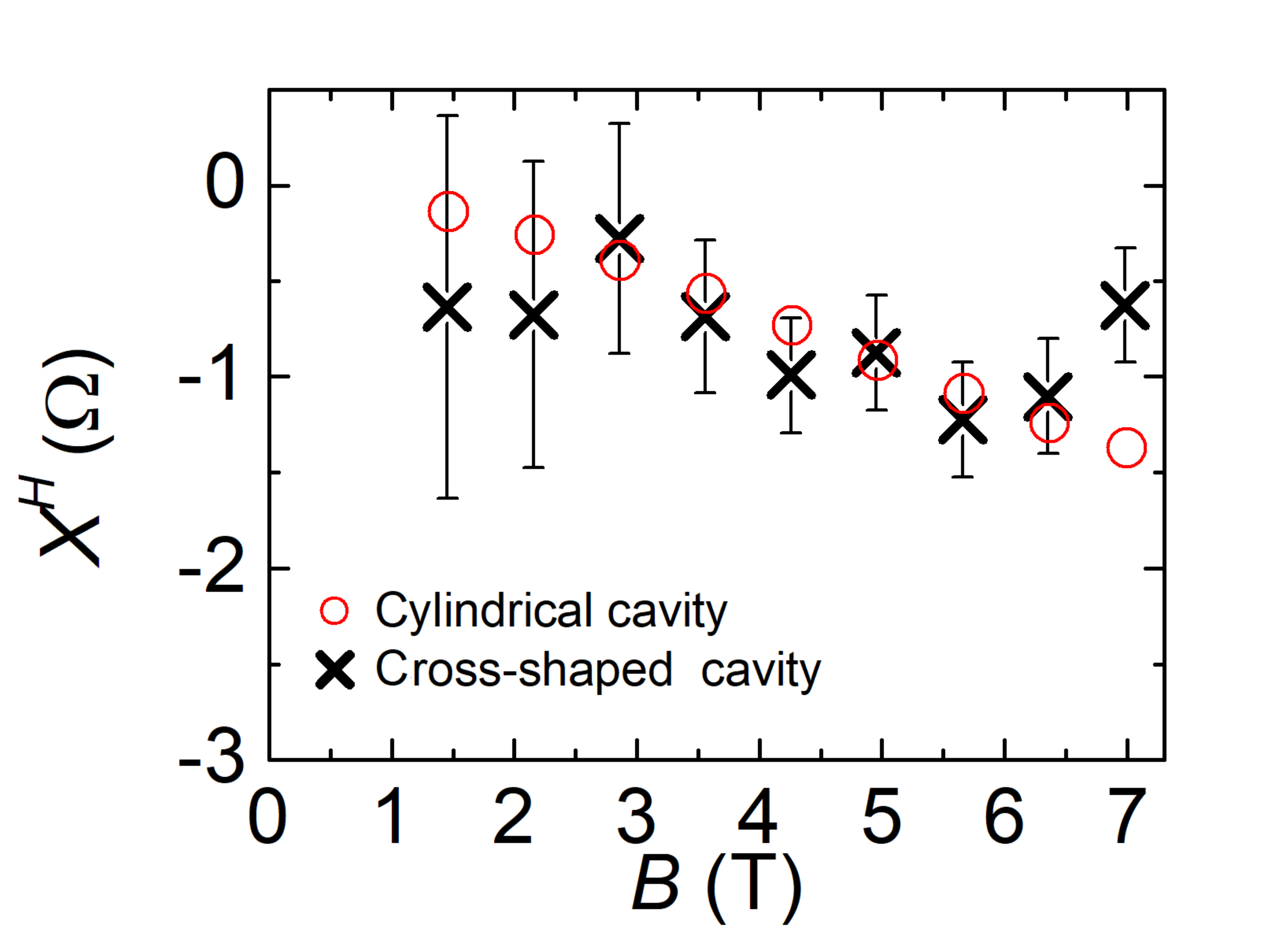}
	\caption{Magnetic fields dependence of $X^H$ of \#S1 at $T=60$ K. 
	Red circles: Data from DC measurements and cylindrical cavity measurements, $X^L\tan\theta_{dc}(T_1,B)$. 
	Cross marks: Data from cross-shaped cavity measurements, $[\Delta(1/2Q_H)(T_1:T_0,B)-G^L \Delta R^L(T_1:T_0,B)]/G^H+X^L\tan\theta_{dc}(T_0,B)$.
	The sign of $X^H$ evaluated from the cross-shaped cavity measurement was determined by the DC measurement. 
	The geometric factor $G^H$ was set to $G^H=6.7\pm 0.8\times 10^{-8}$ $\Omega^{-1}$.}
	\label{GH}
\end{figure}

Then, $G^H$ is determined according to the discussion in the previous section.
The geometric factor $G^H$ was obtained to be $G^H=6.7\pm 0.7\times 10^{-8}$ $\Omega^{-1}$ from Eq.~(\ref{method_DcrossQ}).
The result of this procedure is shown in Fig.~\ref{GH} as a comparative plot of $X^L\tan\theta_{dc}(T_1,B)$ data and $[\Delta(1/2Q)(T_1:T_0,B)-G^L \Delta R^L(T_1:T_0,B)]/G^H+X^L\tan\theta_{dc}(T_0,B)$ data.
The sign of $X^H$ evaluated using the cross-shaped cavity measurement was determined by the DC measurement.

\begin{figure}[htbp]
	\mbox{\includegraphics[keepaspectratio,width=70mm]{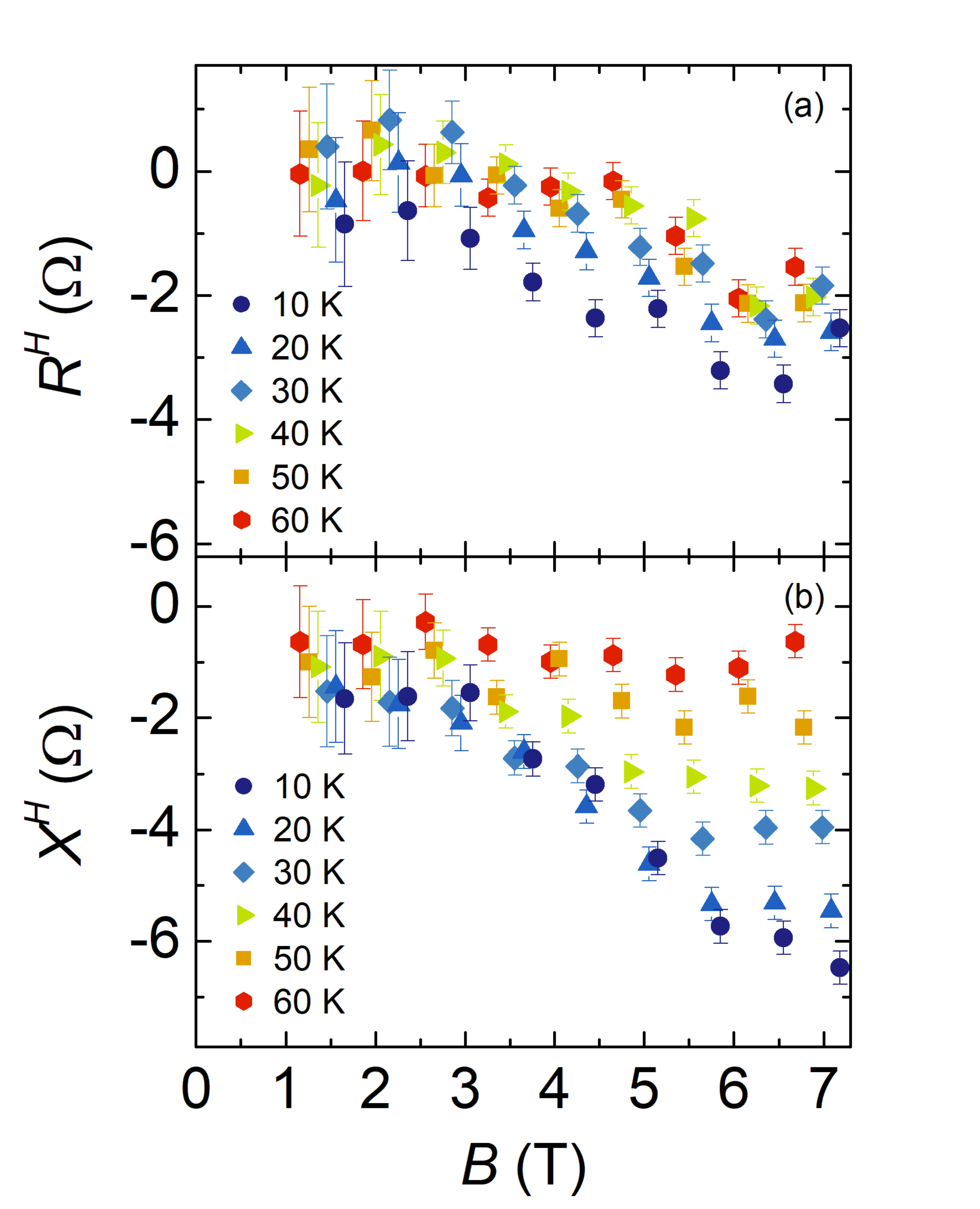}}
	\caption{(a) Transverse surface resistance and (b) transverse surface reactance $X^H$ of \#S1 as a function of the magnetic field at several temperatures.
	The signs  were determined by the DC measurement.}
	\label{RHXH}
\end{figure}

\begin{figure}
	\includegraphics[keepaspectratio,width=70mm]{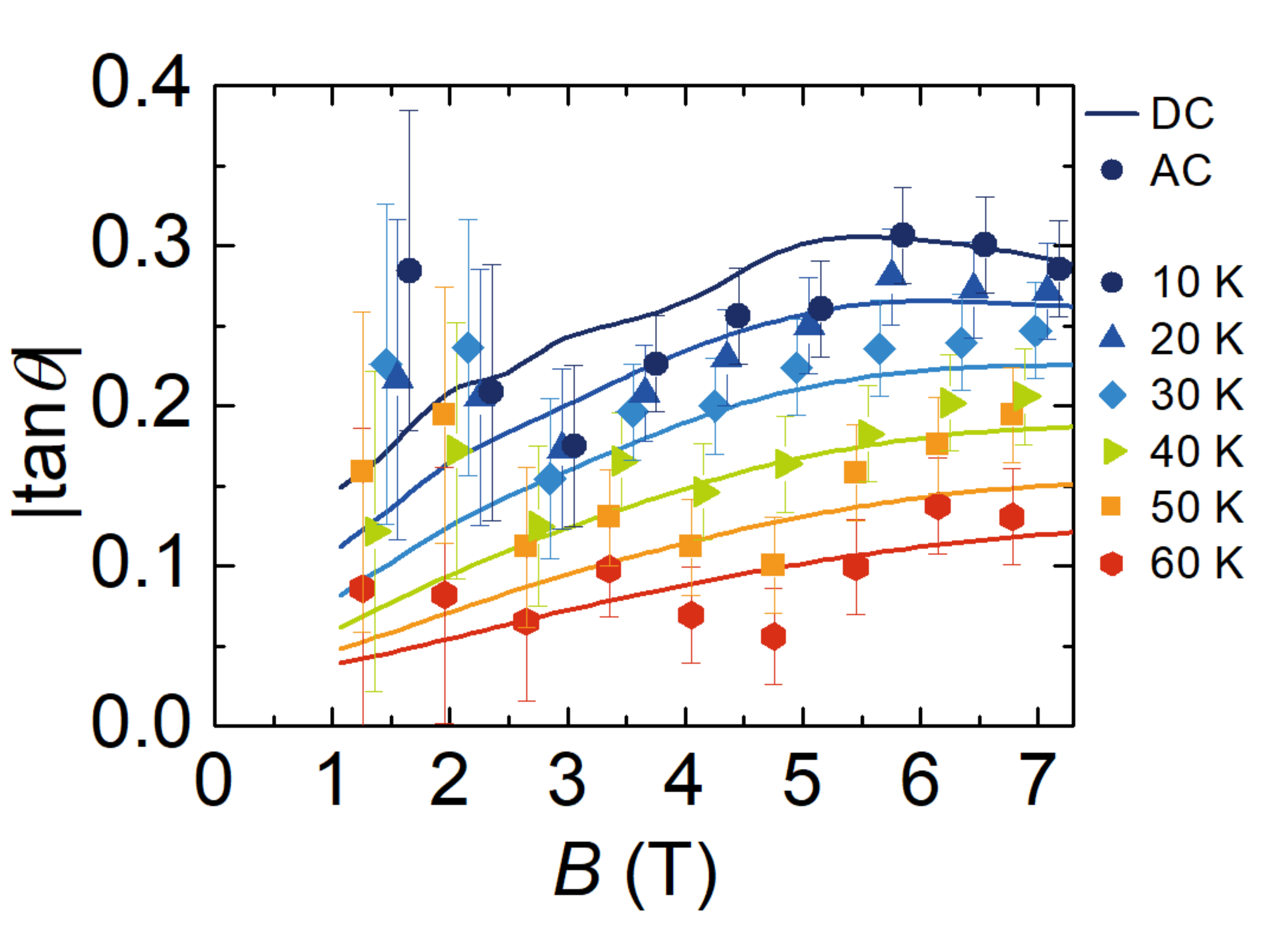}
	\caption{Magnitudes of the tangent of the Hall angle of \#S1 as a function of the magnetic field. 
	Lines are the DC results and marks are the microwave results.}
	\label{tan_result}
\end{figure}

Because geometrical factors were obtained, we evaluate $R^H$ and $X^H$ from Eqs.~(\ref{method_DcrossQ}) and (\ref{method_Dcrossf}).
Fig.~\ref{RHXH} shows the magnetic field dependence of $R^H$ and $X^H$ of Bi from 10 K to 60 K.
The signs of $R^H$ were determined from DC data.
Fig.~\ref{tan_result} shows the tangent of the Hall angle obtained from Eq.~(\ref{method_tan}) together with the DC data.

The magnitude of the microwave Hall angle is on the order of 0.1, and it increases up to $\sim 0.3$ with decreasing temperature. 
Although, at small magnetic fields, the error bars become large because of small signals, the magnitude of the microwave Hall angle coincides with the DC Hall angle.
This suggests that, in most of the data presented above, the condition $\omega\tau\ll 1$ is valid where $\tau$ is the longest life time of the possible carriers in Bi.
Thus, it becomes clear that the Hall angle measurement under cryogenic conditions becomes possible without any complicated tuning mechanisms such as screws, as long as the tangent of the Hall angle is on the order of 0.1; further, our bimodal cavity method can be used to measure the microwave Hall effect on materials in the skin depth region.
This motivates the measurement of the microwave flux-flow Hall effect in superconductors, whose Hall angle $\tan\theta=\omega_0\tau$ is estimated to be  on the order of 0.1 in previous studies,\cite{Tsuchiya2001,Hanaguri1999,Shibata2003,Maeda2007a,Maeda2007,Okada2012,Okada2013a,Okada2013} which will be reported in a different place.\cite{Ogawa2020a}

\section{Conclusion}

We developed a new  microwave Hall effect measurement method for materials in the skin depth region at low temperatures using a cross-shaped bimodal cavity.
We analytically calculated electromagnetic fields in the cross-shaped cavity, and the response of the cavity including the sample, whose property is represented by the surface impedance tensor; further, we constructed the method to obtain the Hall component of the surface impedance tensor in terms of the change in resonance characteristics.
To confirm the validity of the new method, we applied our method to measure the Hall effect in metallic Bi single crystals at low temperatures, and we confirmed that the microwave Hall angles coincide with the DC Hall angle.
Thus, it becomes clear that the Hall angle measurement under cryogenic conditions becomes possible without any complicated tuning mechanisms, and our bimodal cavity method can be used to measure the microwave Hall effect on materials in the skin depth region.
The result opens a new approach to discuss the Hall effect in condensed matter physics such as the microwave flux-flow Hall effect in superconductors. 

\begin{acknowledgments}
We thank Professors S. Fukatsu, Y. Kato, K. Ueno, and Y. Shiomi 
for the fruitful discussions.
This work was partly supported by the Precise Measurement Technology Promotion Foundation (PMTP-F).
We would like to thank Editage (www.editage.com) for English language editing.
\end{acknowledgments}

\section*{Data availability}
The data that support the findings of this study are available from the corresponding author upon reasonable request.

\appendix*
\section{Derivation of formulas used in this study}

In the appendix, we derive  formulas related to the change in the resonant characteristics in the cross-shaped bimodal cavity resonator with the material property of the surface impedance tensor.
First, we calculate the electromagnetic fields in the cross-shaped bimodal cavity, and the resultant change caused by the Hall effect.

\subsection{Surface impedance tensor}

We consider the surface impedance tensor $\tilde{Z}$ containing the Hall effect.\cite{How1990} 
We assume the surface of a highly conductive material represented by $\tilde{\sigma}$ is at $z=$0 in the $xy$ plane.
Suppose the electromagnetic wave is incident from the positive $z$ direction. 

We approximate
$
\nabla\approx\bm{n}\frac{\partial}{\partial n},
$
where $\bm{n}$ is a unit vector perpendicular to the surface of the conductor and toward the inside of the conductor. 
Assuming that the skin depth is expressed as $1/\gamma$, we find the rotations of the electromagnetic fields $\bm{E}$ and $\bm{H}$ as
\begin{subequations}
\begin{eqnarray}
	\nabla\times\bm{E}\approx\bm{n}\times\frac{\partial\bm{E}}{\partial n}=-\gamma\bm{n}\times\bm{E}
\end{eqnarray}
and
\begin{eqnarray}
	\nabla\times\bm{H}\approx\bm{n}\times\frac{\partial\bm{H}}{\partial n}=-\gamma\bm{n}\times\bm{H}.
\end{eqnarray}
\end{subequations}
Thus, from Maxwell's equations,
\begin{subequations}
\begin{eqnarray}\label{eq:maxwell31}
	-\gamma\bm{n}\times\bm{H}\approx\tilde{\sigma}\bm{E}
\end{eqnarray}
and
\begin{eqnarray}\label{eq:maxwell32}
	-\gamma\bm{n}\times\bm{E}\approx i\omega\mu_{0}\bm{H},
\end{eqnarray}
\end{subequations}
where $\mu_0$ is vacuum permeability.

The definition of the surface impedance tensor yields
\begin{equation}\label{eq:maxwell33}
	\bm{E}^{\|}_{n=0}=\tilde{Z}\int^{\infty}_{0}\tilde{\sigma}\bm{E}^{\|}_{n=0}e^{-\gamma n}dn=\frac{1}{\gamma}\tilde{Z}\tilde{\sigma}\bm{E}^{\|}_{n=0},
\end{equation}
where the symbol $\|$ represents that it is parallel to the surface. 
Thus, we obtain 
\begin{eqnarray}\label{appendix_zsigmagannma}
	\tilde{Z}=\gamma\tilde{\sigma}^{-1}.
\end{eqnarray}
As a result, we obtain the  surface impedance tensor  including the Hall effect in the $xy$ plane $\tilde{Z}$ as
\begin{eqnarray}\label{appendix_z}
	\tilde{Z}=
	\frac{\gamma}{\det\tilde{\sigma}}\left(
	\begin{array}{cc}
		1&\tan\theta \\
		-\tan\theta&1\\
	\end{array}
	\right)
	\equiv
	\left(
	\begin{array}{ccc}
		Z^L&Z^{H} \\
		-Z^{H}&Z^L\\
	\end{array}
	\right).
\end{eqnarray}
Thus, the tangent of the Hall angle is expressed by the component of the surface impedance tensor as
\begin{equation}\label{eq:appendix_ZHZ}
\tan\theta=\frac{Z^H}{Z^L}.
\end{equation}

\subsection{Eigenmodes expansion for the electromagnetic field of the cavity}

First, let us review the resonance of an ordinary cavity and derive the equation that relates the resonance characteristics to the surface impedance of the material.
Next, we calculate the resonance characteristics of the bimodal cavity including the material exhibiting the Hall effect.
Therefore, we apply the method developed by Slater,\cite{Slater1946} where we expand the electromagnetic field using eigenmodes.

As discussed in~[44]
, electromagnetic fields $\bs{E}$ and $\bs{H}$ can be expanded by normalized vectors $\bm{E_a}$ and $\bm{H_a}$, which have zero divergence, and $\bm{F_a}$ and $\bm{G_a}$, which have zero rotation as
\begin{subequations}
\begin{equation}
	\bm{E}=\sum_{a}\bm{E}_{a}\int_{V}\bm{E}\cdot\bm{E}_{a}dv+\sum_{a}\bm{F}_{a}\int_{V}\bm{E}\cdot\bm{F}_{a}dv
\end{equation}
and
\begin{equation}
	\bm{H}=\sum_{a}\bm{H}_{a}\int_{V}\bm{H}\cdot\bm{H}_{a}dv+\sum_{a}\bm{G}_{a}\int_{V}\bm{H}\cdot\bm{G}_{a}dv,
\end{equation}
\end{subequations}
where $a$ denotes the index of an eigen mode and $V$ is the volume of the cavity.
These vectors satisfy the following boundary conditions:
\begin{subequations}
\begin{equation}
	\label{eq:BC1}
	\bm{n}\times\bm{E}_a=0\hspace{1.5mm} {\rm on}\hspace{1.5mm} S,
\end{equation}
\begin{equation}
	\label{eq:BC2}
	\bm{n}\cdot\bm{H}_a=0\hspace{1.5mm} {\rm on}\hspace{1.5mm} S,
\end{equation}
\begin{equation}
	\label{eq:BC3}
	\bm{n}\times\bm{H}_a=0\hspace{1.5mm} {\rm on}\hspace{2mm} S^{\prime}
\end{equation}
and
\begin{equation}
	\label{eq:BC4}
	\bm{n}\cdot\bm{E}_a=0\hspace{1.5mm} {\rm on}\hspace{2mm} S^{\prime},
\end{equation}
\end{subequations}
where  $S$ is the conductive surface and $S^{\prime}$ is the insulating surface, respectively, and $\bm{n}$ denotes the unit vector perpendicular to the  surface.
Considering time dependence $e^{-i\omega t}$ and substituting these vectors into Maxwell's equations, we obtain 
\begin{subequations}
\begin{eqnarray}\label{eq:CavityEM1}
i\left(\frac{\omega}{\omega_{a}}-\frac{\omega_{a}}{\omega}\right)\int_{V}\bm{E}\cdot\bm{E}_{a}dv=&&
\frac{i}{\omega\sqrt{\epsilon_{0}\mu_{0}}}\int_{S_0}(\bm{n}\times\bm{E})\cdot\bm{H}_{a}ds\nonumber \\
&&-\frac{1}{\omega_{a}\epsilon_{0}}\int_{S_0}(\bm{n}\times\bm{H})\cdot\bm{E}_{a}ds,\nonumber \\
\end{eqnarray}
\begin{eqnarray}\label{eq:CavityEM2}
i\left(\frac{\omega}{\omega_{a}}-\frac{\omega_{a}}{\omega}\right)\int_{V}\bm{H}\cdot\bm{H}_{a}dv=&&\frac{1}{\omega_{a}\mu_{0}}\int_{S_0}(\bm{n}\times\bm{E})\cdot\bm{H}_{a}ds\nonumber \\
&&+\frac{i}{\omega\sqrt{\epsilon_{0}\mu_{0}}}\int_{S_0}(\bm{n}\times\bm{H})\cdot\bm{E}_{a}ds\nonumber \\
\end{eqnarray}
\end{subequations}
and
\begin{eqnarray}\label{eq:CavityEM3}
\frac{\int_V\bs{E}\cdot\bs{E}_adv}{\int_V\bs{H}\cdot\bs{H}_adv}=i\frac{\omega_a}{\omega}\sqrt{\frac{\mu_0}{\epsilon_0}},
\end{eqnarray}
where $\omega$ denotes the angular frequency of the electromagnetic fields, $\omega_a$ denotes the frequency of the eigen mode $a$, $\epsilon_0$ denotes the vacuum permittivity, and $S_0$ denotes the surface of the cavity resonator system, respectively. 

When we consider the cavity, whose wall is the conductive surface $S$, and the effect of the coupling of the cavity to the outside system on the insulating surface $S^{\prime}$.
From the boundary conditions Eqs.~(\ref{eq:BC1})-(\ref{eq:BC4}), we obtain
\begin{subequations}
\begin{equation}\label{eq:CavityEM4}
\int_{S}(\bm{n}\times\bm{H})\cdot\bm{E}_ads=0,
\end{equation}
\begin{equation}\label{eq:CavityEM5}
\int_{S^{\prime}}(\bm{n}\times\bm{E})\cdot\bm{H}_ads=0
\end{equation}
and
\begin{equation}\label{eq:CavityEM6}
\int_{S^{\prime}}(\bm{n}\times\bm{H})\cdot\bm{E}_ads=\sum_{p}I_{p}v_{a,p},
\end{equation}
\end{subequations}
where $p$ denotes a propagation mode of the waveguide,  $I_p$ denotes the coefficient interpreted as the voltage at plane $S^{\prime}$ in the $p$ waveguide mode set up by $E$, and
 $v_{a,p}$ denotes the coefficient interpreted as the voltage  set up by $E_a$ in the cavity, respectively.\cite{Slater1946} 
Suppose that we apply it in a region where the wave can propagate only in its dominant mode.
Then, we have only one term, which is related to the dominant mode.
Thus, when the cavity is coupled to the waveguide on the cross-section insulating surface $S^{\prime}$, we find that the electromagnetic fields in the cavity satisfy
\begin{subequations}
\begin{eqnarray}\label{eq:CavityEM7}
i\left(\frac{\omega}{\omega_{a}}-\frac{\omega_{a}}{\omega}\right)\int_{V}\bm{E}\cdot\bm{E}_{a}dv
=&&\frac{i}{\omega\sqrt{\epsilon_{0}\mu_{0}}}\int_{S}(\bm{n}\times\bm{E})\cdot\bm{H}_{a}ds\nonumber \\
&&-\frac{1}{\omega_{a}\epsilon_{0}}Iv_{a}
\end{eqnarray}
and
\begin{eqnarray}\label{eq:CavityEM8}
i\left(\frac{\omega}{\omega_{a}}-\frac{\omega_{a}}{\omega}\right)\int_{V}\bm{H}\cdot\bm{H}_{a}dv=&&\frac{1}{\omega_{a}\mu_{0}}\int_{S}(\bm{n}\times\bm{E})\cdot\bm{H}_{a}ds\nonumber \\
&&+\frac{i}{\omega\sqrt{\epsilon_{0}\mu_{0}}}Iv_{a}.
\end{eqnarray}
\end{subequations}
In addition, we obtain
\begin{equation}\label{eq:CavityEM10}
v_a\int_{V}\bm{E}\cdot\bm{E}_{a}dv= V,
\end{equation}
where $V$ is the coefficients interpreted as the voltage at plane $S^{\prime}$ set up by $E$.  \cite{Slater1946}

For the electromagnetic field of the ordinary cylindrical cavity perturbation method, we obtain
\begin{equation}\label{eq:CavityEM11}
\int_S(\bs{n}\times\bs{E})\cdot\bs{H}_ads=-2i\omega\sqrt{\epsilon_0\mu_0}\sum_{i}G_iZ^L_i\int_V\bs{E}\cdot\bs{E}_adv, 
\end{equation}
where index $i$ denotes the effect from the sample or the other elements other than the sample,  $Z^L_i$ denotes the longitudinal components of the surface impedance tensor for each element, and the geometric factor $G_{i}$ is defined as
\begin{eqnarray}\label{eq:CavityEM12}
G_{i}\equiv\frac{\int_{S_{i}}\bm{H}_{a}^{2}ds}{2\mu_{0}\omega_{a}}.
\end{eqnarray}
Thus, together with Eq.~(\ref{eq:CavityEM7}) and Eq.~(\ref{eq:CavityEM10}), we find the input impedance that looks into cavity $Z_0$ as
\begin{eqnarray}\label{eq:CavityEM13}
Z_0=\frac{(v^a)^2}{\omega_a\epsilon_0\left[i(\frac{\omega}{\omega_a}-\frac{\omega_a}{\omega})-2\sum_{i}G_iZ^L_i\right]}.
\end{eqnarray}
Since the input impedance $Z_0$ diverges at the resonance frequency, we obtain the resonance frequency of the cylindrical cavity as
\begin{eqnarray}\label{eq:CavityEM14}
\omega\approx\omega_a(1-i\sum_iG_iZ^L_i),
\end{eqnarray}
where we approximate $1\gg (\sum_iG_iZ^L_i)^2$.
Considering an analogy with the RLC resonance circuit and  the differences of the resonance between with or without the sample expressed by $\Delta_{w/wo}$; we obtain the resonance characteristics of the cylindrical cavity as
\begin{subequations}
\begin{eqnarray}
	\Delta_{w/wo}\left(\frac{1}{2Q}\right)=\frac{1}{2Q}-\frac{1}{2Q_{0}}=GR^L
\end{eqnarray}
and
\begin{eqnarray}
	\Delta_{w/wo}\left(\frac{\omega}{\omega_{0}}\right)=-\frac{\omega-\omega_0}{\omega_0}=GX^L+C,
\end{eqnarray}
\end{subequations}
where we simply write the surface resistance of the sample $R^L_{sample}$ and the surface reactance of the sample $X^L_{sample}$ as $R^L$ and $X^H$, respectively; $\omega$ and $\omega_0$ are the angular resonance frequencies with and without the sample, respectively; $Q$ and $Q_{0}$ are $Q$ factors of the resonance with and without the sample, respectively; $G$ denotes the geometrical constant; and $C$ denotes an experimentally inevitable constant since it is almost impossible to make the cavity resonator exactly the same size when opening and closing it.
In the following section, we extend this discussion to consider the Hall effect in the cross-shaped bimodal cavity.

\subsection{Electromagnetic field in the cross-shaped bimodal cavity}

\begin{figure}[htb]
	\begin{minipage}[t]{0.47\hsize}
		\begin{center}
			\includegraphics[keepaspectratio,width=40mm]{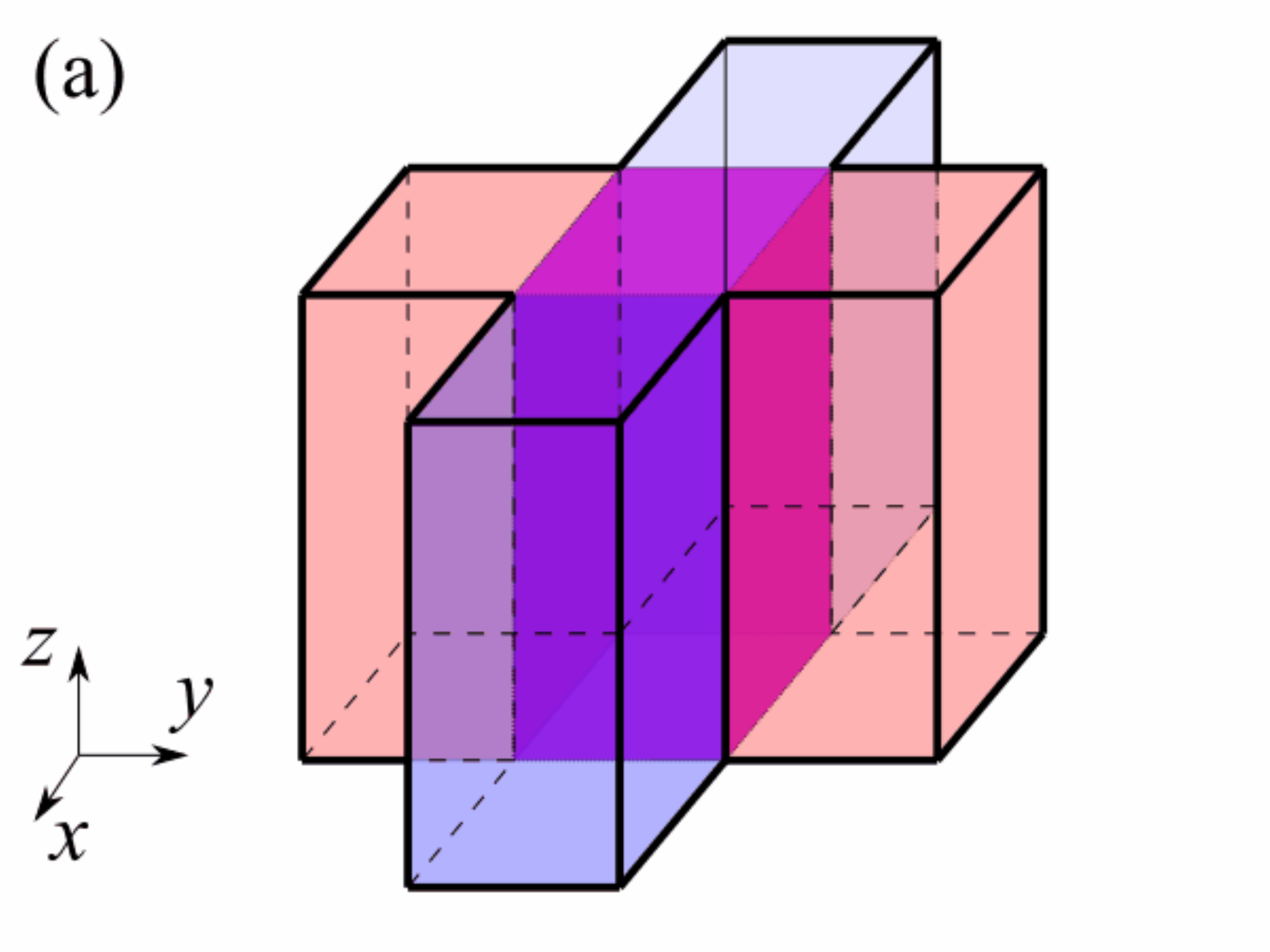}
		\end{center}
	\end{minipage}
	\begin{minipage}{0.02\hsize}
		\hspace{5mm}
	\end{minipage}
	\begin{minipage}[t]{0.47\hsize}
		\begin{center}
			\includegraphics[keepaspectratio,width=40mm]{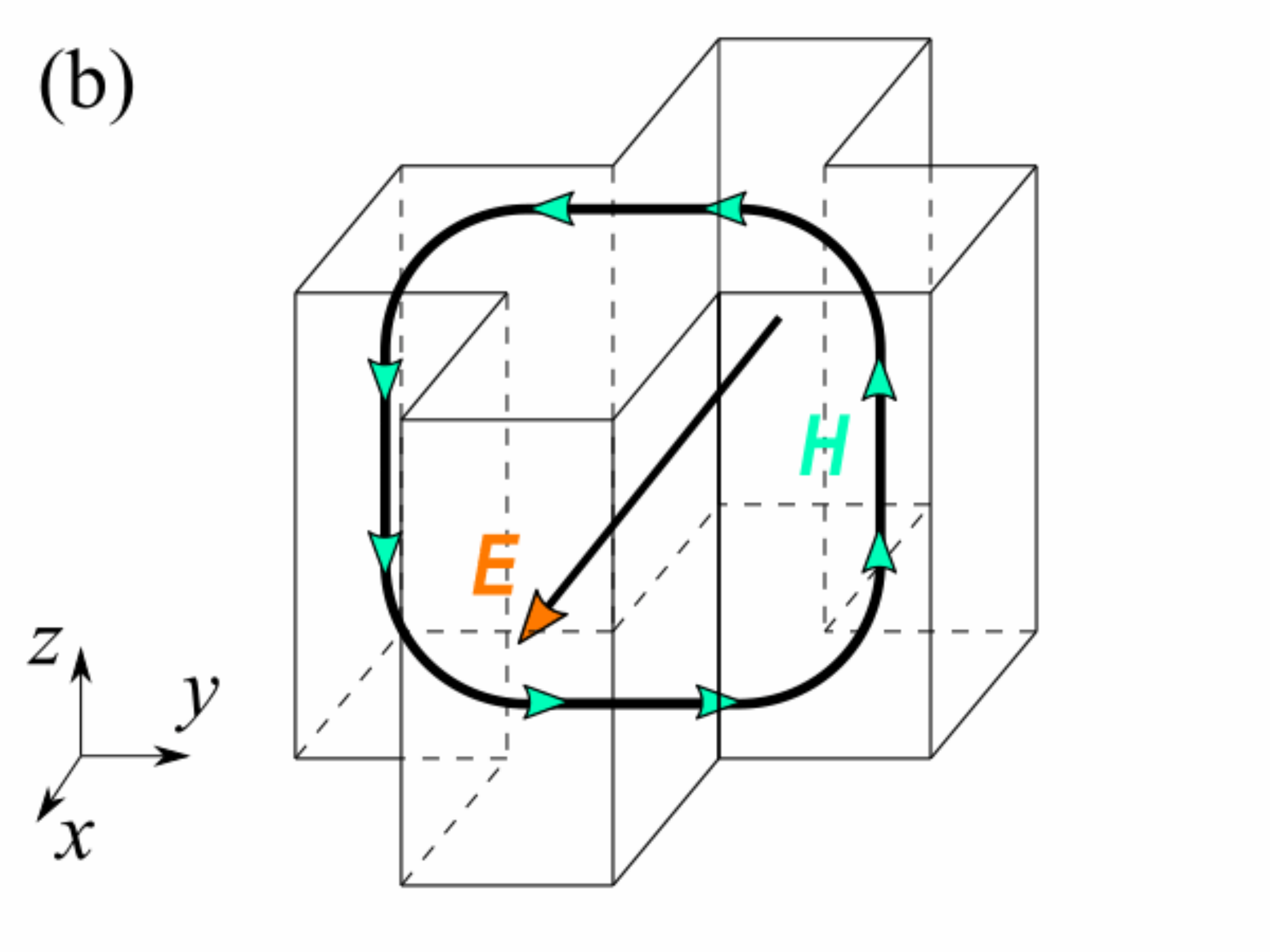}
		\end{center}
	\end{minipage}
		\caption{(a) Cross-shaped bimodal cavity divided into three domains. (1) The center of the cross-shaped bimodal cavity $D_1$ painted in purple, where TE$_{011}$ mode and TE$_{101}$ mode exist and the Hall effect emerge. (2) The part of the cross-shaped bimodal cavity $D_2$  painted in red, where only TE$_{011}$ mode exists. (3)  The part of the cross-shaped bimodal cavity $D_3$ painted in blue where only TE$_{101}$ mode exists. (b) The schematic diagram of TE$_{011}$ mode. The electric field is oriented in the $x$ axis and the magnetic field rotate in the $yz$ plane.}
	\label{Ap_domain}
\end{figure}

In this subsection, we consider a more specific situation, i.e., the electromagnetic field in the cross-shaped bimodal cavity with the sample placed at the center of the cavity.
According to our numerical simulation, in this bimodal cavity, there is the TE$_{011}$ mode where the electric field is oriented in the $x$ axis and the magnetic field rotates in the $yz$ plane; and the TE$_{101}$ mode  where the electric field is oriented in the $y$ axis and the magnetic field rotates in the $xz$ plane.
For simplicity, we assume that the cavity is ideally symmetric and the frequency of both modes are the same, that is, $\omega_{H0}\equiv\omega_{011}=\omega_{101}$. 
In the following calculation, we divide the cavity into three domains: (1) The center of the cross-shaped bimodal cavity $D_1$ where TE$_{011}$ mode and TE$_{101}$ mode exist and the Hall effect emerges. (2) The part of the cross-shaped bimodal cavity in the $y$-axis direction $D_2$, where only the TE$_{011}$ mode exists and we do not consider the Hall effect. (3)  The part of the cross-shaped bimodal cavity in the $x$-axis direction $D_3$, where only the TE$_{101}$ mode exists and  we do not consider the Hall effect (Fig.~\ref{Ap_domain}).

In $D_1$, the magnetic fields are expressed as 
\begin{equation}\label{eq:domain1E}
\bm{H}\approx\bm{H}_{011}\int_{V}\bm{H}\cdot\bm{H}_{011}dv+\bm{H}_{101}\int_{V}\bm{H}\cdot\bm{H}_{101}dv,
\end{equation}
and near the top and bottom of the cavity, the magnetic fields of both modes are expressed as
\begin{subequations}
\begin{eqnarray}\label{eq:domain1_011}
	\bs{H}_{011}\approx H_{011}^y\hat{y}
\end{eqnarray}
and
\begin{eqnarray}\label{eq:domain1_101}
	\bs{H}_{101}\approx H_{101}^x\hat{x}.
\end{eqnarray}
\end{subequations}
From  Eq.~(\ref{eq:maxwell31}) we obtain
\begin{eqnarray}
	\bm{n}\times\bm{E}=\tilde{Z}\bm{H}.
\end{eqnarray}
Thus, together with Eqs.~(\ref{eq:domain1E})-(\ref{eq:domain1_101}), the integrals at the bottom and top of the domain $D_1$ yields
\begin{subequations}
\begin{eqnarray}
	\int_{\delta D_{1}}(\bm{n}\times\bm{E})\cdot\bm{H}_{011}ds
	&&= Z^L\int_{\delta D_{1}}(\bm{H}_{011})^2ds\int_{V}\bm{H}\cdot\bm{H}_{011}dv \nonumber \\
	&&-Z^{H}\int_{\delta D_{1}}H_{101}^xH_{011}^yds\int_{V}\bm{H}\cdot\bm{H}_{101}dv\nonumber \\
\end{eqnarray}
and
\begin{eqnarray}
	\int_{\delta D_{1}}(\bm{n}\times\bm{E})\cdot\bm{H}_{101}ds
	&&=Z^L\int_{\delta D_{1}}(\bm{H}_{101})^2ds\int_{V}\bm{H}\cdot\bm{H}_{101}dv \nonumber \\
	&&+Z^{H}\int_{\delta D_{1}}H_{101}^xH_{011}^yds\int_{V}\bm{H}\cdot\bm{H}_{011}dv, \nonumber \\
\end{eqnarray}
\end{subequations}
where $\delta D_k$ is the conductive surface of  the domain $D_k$ (k=1, 2, 3).

In $D_2$, the magnetic field is expressed as 
\begin{eqnarray}
	\bm{H}\approx\bm{H}_{011}\int_{V}\bm{H}\cdot\bm{H}_{011}dv.
\end{eqnarray}
Thus, the following equations are obtained because the Hall effect is neglected.
\begin{subequations}
\begin{equation}
	\int_{\delta D_{2}}(\bm{n}\times\bm{E})\cdot\bm{H}_{011}ds
	=Z^L\int_{\delta D_{2}}(\bm{H}_{011})^{2}ds\int_{V}\bm{H}\cdot\bm{H}_{011}dv
\end{equation}
and
\begin{equation}
\int_{\delta D_{2}}(\bm{n}\times\bm{E})\cdot\bm{H}_{101}ds=Z^L\int_{\delta D_{2}}(\bm{H}_{101})^{2}ds\int_{V}\bm{H}\cdot\bm{H}_{101}dv=0.
\end{equation}
\end{subequations}
We can perform the same calculation for $D_3$ as $D_2$, and the results are exchanged with respect to the $x$ and $y$ axes.

Combining these calculations in three domains, we find 
\begin{widetext}
\begin{subequations}
\begin{eqnarray}
\sum_i\int_{S_{i}}(\bm{n}\times\bm{E})\cdot\bm{H}_{011}ds&=&\sum_{i}\left(Z^L_{i}\int_{\delta(D_{1}+D_{2}+D_{3}) _{i}}\bm{H}_{011}^{2}ds\int_{V}\bm{H}\cdot\bm{H}_{011}dv-Z^{H}_{i}\int_{\delta (D_{1})_{i}}
H_{101}^xH_{011}^yds\int_{V}\bm{H}\cdot\bm{H}_{101}dv\right)\nonumber \\
&=&-2i\omega_H\sqrt{\epsilon_{0}\mu_{0}}\sum_{i}\left(G^{L}_{i}Z^L_{i}\int_{V}\bm{E}\cdot\bm{E}_{011}dv-G^{H}_{i}Z^{H}_{i}\int_{V}\bm{E}\cdot\bm{E}_{101}dv\right)
\end{eqnarray}
and
\begin{eqnarray}
\sum_i\int_{S_{i}}(\bm{n}\times\bm{E})\cdot\bm{H}_{101}ds&=&\sum_{i}\left(Z^L_{i}\int_{\delta(D_{1}+D_{2}+D_{3}) _{i}}\bm{H}_{101}^{2}ds\int_{V}\bm{H}\cdot\bm{H}_{101}dv+Z^{H}_{i}\int_{\delta(D_{1})_{i}}H_{101}^xH_{011}^yds\int_{V}\bm{H}\cdot\bm{H}_{011}dv\right)\nonumber \\
&=&-2i\omega_H\sqrt{\epsilon_{0}\mu_{0}}\sum_{i}\left(G^{L}_{i}Z^L_{i}\int_{V}\bm{E}\cdot\bm{E}_{101}dv+G^{H}_{i}Z^{H}_{i}\int_{V}\bm{E}\cdot\bm{E}_{011}dv\right),
\end{eqnarray}
\end{subequations}
\end{widetext}
where $i$ denotes the effect from the sample or the other elements other than the sample, $Z^H_i$ denotes the Hall components of the surface impedance tensor for each element, and the geometric factor $G^{L}_{i}$ is defined as
\begin{equation}\label{eq:defGL}
G^{L}_{i}\equiv\frac{\int_{\delta(D_{1}+D_{2}+D_{3}) _{i}}\bm{H}_{101}^{2}ds}{2\mu_{0}\omega_{H0}}
=\frac{\int_{\delta(D_{1}+D_{2}+D_{3}) _{i}}\bm{H}_{011}^{2}ds}{2\mu_{0}\omega_{H0}}
\end{equation}
and $G^H_{i}$ is defined as
\begin{equation}\label{eq:defGH}
G^{H}_{i}\equiv\frac{\int_{\delta (D_{1})_{i}}H_{101}^xH_{011}^yds}{2\mu_{0}\omega_{H0}},
\end{equation}
respectively.
Thus, considering coupling with the two wave guides $\alpha$ and $\beta$, we obtain
\begin{widetext}
\begin{subequations}
\begin{equation}\label{eq:EM31}
i\left(\frac{\omega_H}{\omega_{H0}}-\frac{\omega_{H0}}{\omega_H}\right)\int_{V}\bm{E}\cdot\bm{E}_{011}dv=
2\left(\sum_{i}G^{L}_{i}Z^L_{i}\int_{V}\bm{E}\cdot\bm{E}_{011}dv-\sum_{i}G^{H}_{i}Z^{H}_{i}\int_{V}\bm{E}\cdot\bm{E}_{101}dv\right)
-\frac{1}{\omega_{H0}\epsilon_{0}}I^{(\alpha)}v^{(\alpha)}_{011}-\frac{1}
{\omega_{H0}\epsilon_{0}}I^{(\beta)}v^{(\beta)}_{011}
\end{equation}
and
\begin{equation}\label{eq:EM32}
i\left(\frac{\omega_H}{\omega_{H0}}-\frac{\omega_{H0}}{\omega_H}\right)\int_{V}\bm{E}\cdot\bm{E}_{101}dv=2\left(\sum_{i}G^{L}_{i}Z^L_{i}\int_{V}\bm{E}\cdot\bm{E}_{101}dv+\sum_{i}G^{H}_{i}Z^{H}_{i}\int_{V}\bm{E}\cdot\bm{E}_{011}dv\right)-\frac{1}{\omega_{H0}\epsilon_{0}}I^{(\alpha)}v^{(\alpha)}_{101}-\frac{1}{\omega_{H0}\epsilon_{0}}I^{(\beta)}v^{(\beta)}_{101},
\end{equation}
\end{subequations}
\end{widetext}
where $\alpha$ and $\beta$ mean that these relate to  waveguides $\alpha$ and $\beta$, respectively, and  we consider the dominant propagation mode only.

To go further, we consider the case where the waveguide $\alpha$ is coupled to the TE$_{011}$ mode, whereas, waveguide $\beta$ is coupled to the TE$_{101}$ mode. 
In this case, it is clear that $v_{101}^{(\alpha)}=v_{011}^{(\beta)}=0$.
Thus, Eq.~(\ref{eq:EM31}) and Eq.~(\ref{eq:EM32}) yield
\begin{subequations}
\begin{eqnarray}\label{eq:EM41}
-\frac{I^{(\alpha)}v^{(\alpha)}_{011}}{\omega_{H0}\epsilon_{0}}&=&\left[i\left(\frac{\omega_H}{\omega_{H0}}-\frac{\omega_{H0}}{\omega_H}\right)-2\sum_{i}G^{L}_{i}Z^L_{i}\right]\int_{V}\bm{E}\cdot\bm{E}_{011}dv\nonumber \\
&&+2\sum_{i}G^{H}_{i}Z^{H}_{i}\int_{V}\bm{E}\cdot\bm{E}_{101}dv
\end{eqnarray}
and
\begin{eqnarray}\label{eq:EM42}
-\frac{I^{(\beta)}v^{(\beta)}_{101}}{\omega_{H0}\epsilon_{0}}&=&\left[i\left(\frac{\omega_H}{\omega_{H0}}-\frac{\omega_{H0}}{\omega_H}\right)-2\sum_{i}G^{L}_{i}Z^L_{i}\right]\int_{V}\bm{E}\cdot\bm{E}_{101}dv\nonumber \\
&&-2\sum_{i}G^{H}_{i}Z^{H}_{i}\int_{V}\bm{E}\cdot\bm{E}_{011}dv.
\end{eqnarray}
\end{subequations}
As described in the previous section, we obtain
\begin{subequations}
\begin{eqnarray}\label{eq:int31}
v_{011}^{(\alpha)}\int\bm{E}\cdot\bm{E}_{011}dv= V^{(\alpha)}
\end{eqnarray}
and
\begin{eqnarray}\label{eq:int32}
v_{101}^{(\beta)}\int\bm{E}\cdot\bm{E}_{101}dv= V^{(\beta)},
\end{eqnarray}
\end{subequations}
where $V^{(\alpha)}$ and $V^{(\beta)}$ are the coefficients interpreted as the voltage at the cross-section to the waveguides $\alpha$ and $\beta$, respectively. 
Thus, together with Eq.~(\ref{eq:EM41}) and Eq.~(\ref{eq:EM42}), we obtain
\begin{subequations}
\begin{eqnarray}\label{eq:EM51}
I^{(\alpha)}&=&-\frac{\omega_{H0}\epsilon_{0}}{(v_{011}^{(\alpha)})^2}\left[i\left(\frac{\omega_H}{\omega_{H0}}-\frac{\omega_{H0}}{\omega_H}\right)-2\sum_{i}G^{L}_{i}Z^L_{i}\right]V^{(\alpha)}\nonumber\\
&&-\frac{2\omega_{H0}\epsilon_{0}}{v_{011}^{(\alpha)}v_{101}^{(\beta)}}\sum_{i}G^{H}_{i}Z^{H}_{i}V^{(\beta)}
\end{eqnarray}
and
\begin{eqnarray}\label{eq:EM52}
I^{(\beta)}&=&-\frac{\omega_{H0}\epsilon_{0}}{(v_{101}^{(\beta)})^2}\left[i\left(\frac{\omega_H}{\omega_{H0}}-\frac{\omega_{H0}}{\omega_H}\right)-2\sum_{i}G^{L}_{i}Z^L_{i}\right]V^{(\beta)}\nonumber \\
&&+\frac{2\omega_{H0}\epsilon_{0}}{v_{011}^{(\alpha)}v_{101}^{(\beta)}}\sum_{i}G^{H}_{i}Z^{H}_{i}V^{(\alpha)}.
\end{eqnarray}
\end{subequations}

From Eq.~(\ref{eq:EM51}) and Eq.~(\ref{eq:EM52}), we can see the admittance matrix of the cavity
\be
	\label{eq:admitt}
	\tilde{Y}
	=
	\left(
	\begin{array}{cc}
		Y^{(\alpha\alpha)} &Y^{(\alpha\beta)}\\
		Y^{(\beta\alpha)} & Y^{(\beta\beta)}\\
	\end{array}
	\right),
\ee
expressed as
\begin{subequations}
\begin{equation}
Y^{(\alpha\alpha)}=-\frac{\omega_{H0}\epsilon_{0}}{(v_{101}^{(\alpha)})^2}\left[i\left(\frac{\omega_H}{\omega_{H0}}-\frac{\omega_{H0}}{\omega_H}\right)-2\sum_{i}G^{L}_{i}Z^L_{i}\right],
\end{equation}
\begin{equation}
Y^{(\alpha\beta)}=-\frac{2\omega_{H0}\epsilon_{0}}{v_{011}^{(\alpha)}v_{101}^{(\beta)}}\sum_{i}G^{H}_{i}Z^{H}_{i},
\end{equation}
\begin{equation}
Y^{(\beta\alpha)}=\frac{2\omega_{H0}\epsilon_{0}}{v_{011}^{(\alpha)}v_{101}^{(\beta)}}\sum_{i}G^{H}_{i}Z^{H}_{i}
\end{equation}
and
\begin{equation}
Y^{(\beta\beta)}=-\frac{\omega_{H0}\epsilon_{0}}{(v_{101}^{(\beta)})^2}\left[i\left(\frac{\omega_H}{\omega_{H0}}-\frac{\omega_{H0}}{\omega_H}\right)-2\sum_{i}G^{L}_{i}Z^L_{i}\right].
\end{equation}
\end{subequations}
In the following, we assume that $v_{011}^{(\alpha)}=v_{101}^{(\beta)}=1$ for simplicity.
From the assumption, we can rewrite the elements of the admittance matrix as
\begin{subequations}
\begin{equation}
Y^{(\alpha\alpha)}=Y^{(\beta\beta)}=K_{1}
\end{equation}
and
\begin{equation}
-Y^{(\alpha\beta)}=Y^{(\beta\alpha)}=K_{2},
\end{equation}
\end{subequations}
where $K_1$ is defined as
\begin{equation}\label{eq_K1}
K_{1}\equiv\omega_{H0}\epsilon_{0}\left[-i\left(\frac{\omega_H}{\omega_{H0}}-\frac{\omega_{H0}}{\omega_H}\right)+2\sum_{i}G^{L}_{i}Z^L_{i}\right]
\end{equation}
and $K_2$ is defined as
\begin{equation}\label{eq_K2}
K_{2}\equiv2\omega_{H0}\epsilon_{0}\sum_{i}G^{H}_{i}Z^{H}_{i}.
\end{equation}

\subsection{Resonance characteristics of the cross-shaped bimodal cavity}

At the resonance frequency, the input impedance of the cavity diverges and the determinant of the admittance matrix becomes zero. 
Thus, from the determinant of $\tilde{Y}$ in Eq.~(\ref{eq:admitt}), we find 
\begin{eqnarray}
K_{1}^{2}+K_{2}^{2}=0.
\end{eqnarray}
Solving this equation yields
\begin{eqnarray}
\omega_H&\approx&\omega_{H0}(1+\sum_{i}G^{L}_{i}X^L_{i}\mp\sum_{i}G^{H}_{i}R^{H}_{i})\nonumber \\
&&-i\omega_{H0}(\sum_{i}G^{L}_{i}R^L_{i}\pm\sum_{i}G^{H}_{i}X^{H}_{i}),
\end{eqnarray}
where we approximate $1\gg (\sum_iG^L_iZ^L_i\pm i\sum_iG^H_iZ^H_i)^2$.
By considering an analogy with the RLC resonance circuit,  we obtain the resonance characteristics as
\begin{subequations}
\begin{eqnarray}
\frac{1}{2Q_H}=\sum_{i}G^{L}_{i}R^L_{i}\pm\sum_{i}G^{H}_{i}X^{H}_{i}
\end{eqnarray}
and
\begin{eqnarray}
-\frac{\omega_H-\omega_{H0}}{\omega_{H0}}=\sum_{i}G^{L}_{i}X^L_{i}\mp\sum_{i}G^{H}_{i}R^{H}_{i},
\end{eqnarray}
\end{subequations}
where $Q_H$ and $\omega_H=2\pi f_H$ are the quality factor and resonance angular frequency of the cross-shaped bimodal cavity in the Hall direction, respectively.
Representing  the influence to the $Q$ factor and resonance angular frequency from the other elements other than the sample as $D_1$ and $D_2$, respectively, we obtain 
\begin{subequations}
\begin{equation}\label{eq:appendix11}
\frac{1}{2Q_H}=G^{L}R^L\pm G^HX^H+D_1
\end{equation}
and
\begin{equation}
-\frac{\omega_{H}-\omega_{H0}}{\omega_{H0}}=G^{L}X^L\mp G^HR^H+D_2,
\end{equation}
\end{subequations}
where we abbreviate the subscripts "sample" (e.g., $R_L=R^L_{sample}$).

Considering the case where an ordinary sample is in a resonator made of perfect conductor namely $D_1=0$, if the sign of the Hall term in Eq.~(\ref{eq:appendix11}) is negative, $Q$ diverges for some $R$ and $X^H$ despite the loss of the sample. Therefore, we obtain
\begin{subequations}
\begin{equation}
\frac{1}{2Q_H}=G^{L}R^L+ G^H|X^H|+D_1
\end{equation}
and
\begin{equation}
-\frac{\omega_{H}-\omega_{H0}}{\omega_{H0}}=G^{L}X^L- G^H|R^H|+D_2.
\end{equation}
\end{subequations}
Considering the differences of the resonance between with or without the sample, expressed by $\Delta_{w/wo}$, we rewrite the above equations as
\begin{subequations}
\begin{equation}\label{AppRes1}
\Delta_{w/wo}\left(\frac{1}{2Q_H}\right)\equiv\frac{1}{2Q_H}-\frac{1}{2Q_{H0}}=G^LR^L+G^H|X^H|
\end{equation}
and
\begin{equation}\label{AppRes2}
\Delta_{w/wo}\left(\frac{\omega_H}{\omega_{H0}}\right)\equiv-\frac{\omega_{H}-\omega_{H0}}{\omega_{H0}}=G^LX^L-G^H|R^H|+D,
\end{equation}
\end{subequations}
where $Q_{H0}$ denotes the $Q$ factor of the resonance without the sample, $D$ denotes an experimentally inevitable constant because it is almost impossible to make the cavity resonator exactly the same size when opening and closing it.
As mentioned in the Method, these formulas can be interpreted in terms of the adiabatic theorem which states that the shift in the complex  frequency is equal to the change in the complex energy of the cavity. 
Thus, we can consider these formulas as a natural extension of the equations in the ordinary cavity perturbation technique.

\end{document}